\pdfoutput=1

\documentclass[showpacs,nofootinbib,showkeys,eqsecnum,prd,aps,preprint]{revtex4-2}

\usepackage[english]{babel}
\usepackage{comment}

\usepackage{amssymb}
\usepackage{amsmath}
\usepackage{amsfonts}
\usepackage{latexsym}
\usepackage{mathrsfs}
\usepackage{bm}
\usepackage{tensor}
\usepackage{hyperref}
\usepackage{graphicx}

\def\Sp{\mathop{\rm Sp}\nolimits\,}

\def\dac{\displaystyle\frac}
\def\dil{\displaystyle\int\limits}

\usepackage{amsthm}

\newtheorem{defin}{Definition}[section]
\newtheorem{demo}{Proposition}[section]

\usepackage{diagbox}

\def\pa{\partial}

\def\BFC{{\bf C}}

\def\Sp{\mathop{\rm Sp}\nolimits\,}

\def\dac{\displaystyle\frac}

\def\dil{\displaystyle\int\limits}
\def\{{\lbrace}
\def\}{\rbrace}

\def\Or{{\rm O}}

\begin{document}

\title{Semiclassical approach to the nonlocal nonlinear Schr\"{o}dinger equation with a non-Hermitian term}

\author{Anton E. Kulagin}
\email{aek8@tpu.ru}
\affiliation{Tomsk Polytechnic University, 30 Lenina av., 634050 Tomsk, Russia}

\author{Alexander V. Shapovalov}
\email{shpv@mail.tsu.ru}
\affiliation{Department of Theoretical Physics, Tomsk State University, 1 Novosobornaya Sq., 634050 Tomsk, Russia}
\affiliation{Laboratory for Theoretical Cosmology, International Centre of Gravity and Cosmos, Tomsk State University of Control Systems and Radioelectronics, 40 Lenina av., 634050 Tomsk, Russia}

\begin{abstract}
The nonlinear Sch\"{o}dinger equation (NLSE) with a non-Hermitian term is the model for various phenomena in nonlinear open quantum systems. We deal with the Cauchy problem for the nonlocal generalization of multidimensional NLSE with a non-Hermitian term. Using the ideas of the Maslov method, we propose the method of constructing asymptotic solutions to this equation within the framework of semiclassically concentrated states. The semiclassical nonlinear evolution operator and symmetry operators for the leading term of asymptotics are derived. Our approach is based on the solutions of the auxiliary dynamical system that effectively linearize the problem under certain algebraic conditions. The formalism proposed is illustrated with the specific example of the NLSE with a non-Hermitian term that is the model of an atom laser. The analytical asymptotic solution to the Cauchy problem is obtained explicitly for this example.\\
\end{abstract}



\maketitle

\section{Introduction}
\label{sec:int}

A great variety of nonlinear phenomena in  matter-wave and optical media
is modelled based on generalizations and modifications of the nonlinear Schr\"odinger equation (NLSE). A significant share of theoretical research based on NLSE is associated with nonlinear optics \cite{lederer2008}, the physics of Bose--Einstein condensates (BEC) \cite{pitaevskii1999,leggett2001}, the dynamics of quantum vortices \cite{saffman1993,fetter01}, and other areas of nonlinear physics. In the BEC theory based on the mean-field approximation, the NLSE, termed the Gross--Pitaevskii equation (GPE) \cite{pitaevskii1999}, is the base model equation. Mathematically, it is considered in multidimensional space-time with the variable coefficients responsible for the  external fields of the traps confining the condensate in  some area. In nonlinear optics, the NLSE describes optical solitons \cite{hasegawa1973,mollenauer1980,zakharov72}, which represent spatially localized perturbations of an electromagnetic field steadily propagating in a nonlinear medium.


These models usually describe conservative systems isolated from the environment, in which dissipative phenomena are not taken into account. However, in real conditions, quantum systems interact with the environment. In many-particle quantum systems, this interaction is dissipative, weakening the coherent effects, and, accordingly, blurring the manifestation of pronounced quantum properties. On the other hand, the combination of dissipative effects with quantum ones gives rise to new scenarios of behavior in nonlinear systems. This encourages the interest in the study of dissipative phenomena in nonlinear quantum systems, primarily BEC and nonlinear optical ones that has been the subject of detailed studies in many publications. Some examples below give an idea of these studies.

%
%
%


In \cite{arecchi2000} the use of BEC atoms in an atom laser is discussed. This problem is directly related to the interaction of the condensate with noncondensed trapped atoms which is a dissipative process. The theoretical description of this system is constructed in the mean field approximation and GPE with additional dissipative terms. After some simplifications, we obtain a closed description in terms of the complex Ginzburg--Landau  equation (GLE) \cite{aranson2002}, which formally has the NLSE form with a non-Hermitian operator. In the review \cite{brazhnyi2004}, in particular, the impact on the BEC parameters by Feshbach resonance, which leads to the generation of trains of solitons, was studied. This problem was considered in terms of a dissipative one-dimensional GPE with a time-dependent complex addition to the potential. The paper \cite{abdullaev2010} considers the dynamics of nonlinear waves in periodic complex PT-potentials, when the NLSE becomes invariant under the parity and time-reversal symmetry. A special kind of non-equilibrium stationary states was introduced and studied in \cite{sels2020} based on the GPE with a stochastic noise term, which is considered by adding terms that violate the hermiticity of the GPE operator. The paper emphasizes that the states under study and the resulting specific phase transitions in BEC are possible only in the presence of dissipation. The dissipative NLSE also arise in the description of solitons in nonlinear media such as the cavity of the mode-locked lasers (the so-called Haus master equation \cite{haus1984} that is the (1+1)-NLSE with a non-Hermitian term) and related models \cite{aleksic21} including multidimensional \cite{aleksic20} and nonlocal \cite{aleksic14} ones.

Most of the studies devoted to the mentioned nonlinear systems, including the BEC, deal with the local form of the NLSE since it is simpler for the mathematical analysis especially for the numerical one. While such simplification is reasonable for short range interactions, it can not be used for long range interactions. The example of the last ones is dipole-dipole interaction \cite{baranov2008,malomed2009} that significantly affects the dynamics of the BEC \cite{klaus2022,zhao2021}. In \cite{curtis2012}, the author studied the influence of the nonlocal interaction of condensate particles and an external periodic field on the BEC dynamics in the framework of the mean field theory and nonlocal generalized GPE without dissipative terms. Therefore, it is of interest to study the nonlocal NLSE as a more general problem since the well studied local form can be treated as the limiting case of the nonlocal one in a manner.

Using the nonlocal form of the model equations, we apply the semiclassical formalism to the problem under consideration. The semiclassical approximation is widely used for linear equations of quantum mechanics. Some modern semiclassical approaches based on ideas of the Maslov method \cite{Maslov2} were also applied to some nonlinear problems (see, e.g., \cite{dobrokhotov2022,pereskokov2022}). In \cite{shapovalov:BTS1,sym2020,kulagin2021}, the formalism of semiclassical asymptotics for a generalized nonlocal GPE in a special class of trajectory concentrated functions is developed that corresponds to closed quantum systems. In \cite{fkppshap2018,shapkul21,shapkul22}, this formalism was applied to kinetic reaction-diffusion equations that correspond to open classical systems. The conception of this work is to combine the ideas of those approaches to solve the nonlinear problem corresponding to open quantum systems.

In this paper, following \cite{shapovalov:BTS1,sym2020,fkppshap2018,shapkul21}, we extend the method of semiclassical asymptotics as applied to a generalized nonlocal NLSE with a non-Hermitian term that is responsible for the dissipation. In particular cases, the equation considered in the paper transforms into a complex nonlocal GLE \cite{aranson2002}, as well as into a nonlocal NLSE with a complex potential \cite{nath2017,abdullaev2010}. The general formalism is illustrated by an example.

The paper is organized as follows. In Section \ref{sec:nlse}, the original nonlinear problem is posed. In Section \ref{sec:sred}, we introduce the moment of the desired solution and give some additional notations. In Section \ref{sec:class}, we explain what we mean by the semiclassically concentrated states, derive the classical equations corresponding to the nonlinear quantum problem, and introduce the class of functions where the asymptotic solutions are sought. In Section \ref{sec:mom}, we deal with the auxiliary dynamical system that allows us to proceed to the linear partial differential equation associated with the original nonlinear problem. In Section \ref{sec:alse}, we derive this linear partial differential equation and construct the leading term of the asymptotic solution to the Cauchy problem for the original nonlinear equation under some algebraic conditions. The explicit analytical form of the semiclassical nonlinear evolution operator is given. In Section \ref{sec:sym}, we discuss the semiclassical symmetries for the problem under consideration. Section \ref{sec:example} provides the example for the presented formalism. Here, we apply our method to the specific NLSE that is the model of an atom laser. Analytical asymptotic solutions are obtained for this equation. In Section \ref{sec:con}, we conclude with some remarks.

%
%
%
%
%

\section{Nonlocal NLSE with a non-Hermitian term}
\label{sec:nlse}

Let us write the non-stationary nonlocal NLSE with a non-Hermitian term as follows:
\begin{equation}
\begin{array}{l}
\bigg\{ -i\hbar\pa_t + H(\hat{z},t)[\Psi]-i\hbar \Lambda \breve{H}(\hat{z},t)[\Psi] \bigg\}\Psi(\vec{x},t)=0, \cr \cr
H(\hat{z},t)[\Psi]=V(\hat{z},t)+\varkappa\dil_{{\mathbb{R}}^n}d\vec{y}\,\Psi^{*}(\vec{y},t)W(\hat{z},\hat{w},t)\Psi(\vec{y},t), \cr
\breve{H}(\hat{z},t)[\Psi]=\breve{V}(\hat{z},t)+\varkappa\dil_{{\mathbb{R}}^n}d\vec{y}\,\Psi^{*}(\vec{y},t)\breve{W}(\hat{z},\hat{w},t)\Psi(\vec{y},t).
\end{array}
\label{hartree1}
\end{equation}
Here,  $t\in{\mathbb{R}}^1 $, $\vec{x}\in {\mathbb{R}}^n$, $\Lambda$ and $\varkappa$ are real non-Hermiticity and nonlinearity parameters, respectively, $\hbar$ acts as a formal small asymptotic parameter \cite{Maslov2}, and symbol $({}^{*})$ denotes the complex conjugation. The operators $V(\hat{z},t)$, $\breve{V}(\hat{z},t)$, $W(\hat{z},\hat{w},t)$, and $\breve{W}(\hat{z},\hat{w},t)$  in \eqref{hartree1} depend on noncommuting operators $\hat{z}=(\hat{\vec{p}},\vec{x})$, $\hat{w}=(\hat{\vec{p}}_y,\vec{y})$,  $\hat{\vec{p}}=-i\hbar\dac{\partial}{\partial\vec{x}}=-i\hbar\nabla$, $\hat{\vec{p}}_y=-i\hbar\dac{\partial}{\partial\vec{y}}$, $\vec{y}\in {\mathbb{R}}^n$. Note we put the arrows only for $n$-dimensional vectors. We do not puts ones for $z$ that is $2n$-dimensional.

We consider solutions $\Psi$ to the equation \eqref{hartree1} that belong to the Schwartz space $\mathbb{S}$ with respect to $\vec{x}$  and deal with the $L_2({\mathbb{R}_x^n})$ scalar product
  \begin{align}
\langle \Phi|\Psi \rangle(t)= \displaystyle\int\limits_{{\mathbb{R}}^n}
d\vec{x}\,\Phi^{*}(\vec{x},t)\Psi(\vec{x},t).
\label{scal1}
\end{align}

The operators $\hat{z},\hat{w}$ satisfy the following commutation relations:
 \begin{equation}
 [\hat{z}_k,\hat{z}_j]=i\hbar J_{kj},\quad j,k=1,\dots , 2n,
 \label{comm1}
  \end{equation}
where $J=  \left(
    \begin{array}{cc}
      0 & -I_{n} \\
      I_{n} &0 \\
    \end{array}
  \right)
  $
  is the $2n\times 2n$ symplectic identity matrix, $I_{n}$ is the $n\times n$ identity matrix. The commutators and anticommutators of operators $\hat{A}$ and $\hat{B}$ are denoted by $[\hat{A}, \hat{B}]=\hat{A}\hat{B}-\hat{B}\hat{A}$ and $[\hat{A}, \hat{B}]_{+}=\hat{A}\hat{B}+\hat{B}\hat{A}$, respectively.
  The scalar products of vectors from ${\mathbb{R}}^{n}$ and ${\mathbb{R}}^{2n}$ are denoted by $\langle a,b \rangle=\sum_{j}a_j b_j$.

Let us introduce the set $\mathcal{S}$ of functions $A(z,t,\hbar)$=$A(\vec{p},\vec{x},t,\hbar)$ that satisfy the following conditions for every fixed $t\geq 0$:\\
1) $A(\vec{p},\vec{x},t,\hbar)\in C^{\infty}$ with respect to $\vec{p}$ and $\vec{x}$;\\
2) $A(\vec{p},\vec{x},t,\hbar)$ and all its derivatives grow not faster that polynomials of $|\vec{p}|$ and $|\vec{x}|$ as $|\vec{p}|,|\vec{x}|\to \infty$;\\
3) $A(\vec{p},\vec{x},t,\hbar)$ regularly depends on the parameter $\hbar$ in a neighborhood of $\hbar=0$. \\

\begin{defin}
A pseudo-differential Weyl-ordered operator is an operator $\hat{A}=A(\hat{z},t,\hbar)=A(\hat{\vec{p}},\vec{x},t,\hbar)$ that is defined by \cite{Maslov1}
\begin{equation}
A(\hat{\vec{p}},\vec{x},t,\hbar)\Phi(\vec{x},t,\hbar)=\dac{1}{(2\pi\hbar)^n}\dil_{{\mathbb{R}}^{2n}}d\vec{p}d\vec{y} \exp\Big(\dac{i}{\hbar}\langle \vec{p},\vec{x}-\vec{y}\rangle\Big)A\Big(\vec{p},\dac{\vec{x}+\vec{y}}{2},t,\hbar\Big)\Phi(\vec{y},t,\hbar),
\label{pseud1}
\end{equation}
where $A(\vec{p},\vec{x},t,\hbar)\in{\mathcal{S}}$ and $\Psi(\vec{x},t,\hbar)\in{\mathbb{S}}$ for fixed $t$, $\hbar$.

The function $A(z,t,\hbar)$ in \eqref{pseud1} is termed the Weyl symbol of the operator $\hat{A}=A(\hat{z},t,\hbar)$.

We denote by $\mathcal{A}$ the set of pseudo-differential operators defined above.
\end{defin}

The operators $V(\hat{z},t)$, $\breve{V}(\hat{z},t)$, $W(\hat{z},\hat{w},t)$, and $\breve{W}(\hat{z},\hat{w},t)$  in \eqref{hartree1} belong to $\mathcal{A}$. The functions $V(z,t)$, $\breve{V}(z,t)$, $W(z,w,t)$, and $\breve{W}(z,w,t)$ are their Weyl symbols, respectively. These operators are Hermitian with respect to the scalar product \eqref{scal1}.

Whenever no confusion arises, we will simplify our notation. We will drop the explicit dependence on $\hbar$ in the functions and indicate it where appropriate, in particular, in  solutions $\Psi$ to equation \eqref{hartree1}, $\Psi(\vec{x},t,\hbar)=\Psi(\vec{x},t)$.

\section{Expectation of an operator over functions from $\mathcal{S}$}
\label{sec:sred}

A non-Hermitian term in the operator of the equation \eqref{hartree1} results in that $L_2$-norm of the solution $\Psi(\vec{x},t,\hbar)$,  $\|\Psi\|^2(t,\hbar)=\langle\Psi|\Psi\rangle(t,\hbar)$,  does not conserve. Hereinafter, $||\cdot||$ stands for the $L_2$-norm. Let us denote
\begin{equation}
\sigma_{\Psi}(t,\hbar)=\|\Psi\|^2(t,\hbar)
 \label{sig1}
 \end{equation}
and derive the evolution equation for $\sigma_{\Psi}(t,\hbar)$.
From \eqref{hartree1}, taking into account the Hermiticity of operators  $H(\hat{z},t)[\Psi]$ and $\breve{H}(\hat{z},t)[\Psi] $, one gets
 \begin{equation}
\dot{\sigma}_{\Psi}(t,\hbar)=-2\Lambda \displaystyle\int\limits_{{\mathbb{R}}^n}
d\vec{x}\,\Psi^{*}(\vec{x},t;\hbar) \breve{H}(\hat{z},t)[\Psi]\Psi(\vec{x},t;\hbar)=
-2\Lambda\langle\Psi|\breve{H}[\Psi]|\Psi \rangle,
 \label{sig2}
 \end{equation}
where $\dot{\sigma}_{\Psi}=d\sigma/dt$, and $\breve{H}(\hat{z},t)[\Psi]$ is given by \eqref{hartree1}. We will consider the solutions with $\sigma_{\Psi}(t,\hbar)=\Or(1)$ as $\hbar\to 0$. Actually, this condition depends on the definition of the nonlinearity coefficient $\varkappa$. We will limit our consideration to the case $\varkappa=\Or(1)$ since it is of the greatest interest from the physical point of view. It corresponds to the situation when the linear and nonlinear parts of the equation \eqref{hartree1} are comparable. In this case, regularly perturbation theory in nonlinearity parameter does not yield qualitative result.

For the operator $\hat{A}(t) \in\mathcal{A}$ determined by its Weyl symbol $\hat{A}(t)=A(\hat{z},t)$, we define the expectation for the solution $\Psi(\vec{x}, t,\hbar)$ 
to \eqref{hartree1} by the following relation:
 \begin{equation}
\langle \hat{A}(t) \rangle_\Psi= \frac{1}{\sigma_\Psi(t,\hbar)}\langle\Psi| \hat{A}(t)|\Psi\rangle=\frac{1}{\sigma_\Psi(t,\hbar)}
\int\limits_{{\mathbb{R}}^n}d\vec{x}\,\Psi^{*}(\vec{x},t,\hbar)A(t)(\hat{z},t)\Psi(\vec{x},t,\hbar).
 \label{mean1}
 \end{equation}

The evolution equation for the expectation $\langle \hat{A}(t) \rangle_\Psi$ on solutions to \eqref{hartree1} reads
\begin{align}
&\displaystyle\frac{\partial}{\partial t}\langle\hat{A}(t)\rangle_\Psi=-\frac{\partial\log\sigma_\Psi(t,\hbar)}{\partial t} \langle\hat{A}(t)\rangle_\Psi+\bigg\langle\displaystyle\frac{\partial \hat{A}(t)}{\partial t} \bigg\rangle_\Psi +\frac{i}{\hbar}\big\langle \big[H(\hat{z},t)[\Psi],\hat{A}(t)\big]
 \big\rangle_\Psi-\cr
 &-\Lambda \big\langle \big[\breve{H}(\hat{z},t)[\Psi],\hat{A}(t)\big]_{+}
 \big\rangle_\Psi= -\frac{\partial\log\sigma_\Psi(t,\hbar)}{\partial t} \langle\hat{A}(t)\rangle_\Psi+\bigg\langle\displaystyle\frac{\partial \hat{A}(t)}{\partial t} \bigg\rangle_\Psi+  \cr
 & + \dac{i}{\hbar} \big\langle [V(\hat{z},t),A(\hat{z},t)]\big\rangle_\Psi - \Lambda\big\langle [\breve{V}(\hat{z},t),A(\hat{z},t)]_{+}\big\rangle_\Psi + \cr
&+\varkappa \bigg\langle\dil_{{\mathbb{R}}^n}d\vec{y}\, \Psi^{*} \Big(\dac{i}{\hbar}[W(\hat{z},\hat{w},t), A(\hat{z},t)]-\Lambda [\breve{W}(\hat{z},\hat{w},t), A(\hat{z},t)]_{+} \Big)\Psi(\vec{y},t)\bigg\rangle_\Psi.
  \label{mean2}
\end{align}

The equations \eqref{mean2} can be significantly simplified in the semiclassical approximation and such simplified equations will determine the semiclassical evolution of the solutions to \eqref{hartree1}. In the next Section, we will clarify how we interpret the semiclassical limit within the framework of our approach.

\section{Class of semiclassically concentrated functions}
\label{sec:class}

\begin{defin}
The function $\Psi(\vec{x},t,\hbar)$ belongs to a class ${\mathcal{T}}^t_\hbar(Z(t),\sigma(t))$ of functions semiclassically concentrated in a neighborhood of a trajectory $z=Z(t)$ with a weight of $\sigma(t)$ if for any operator $\hat{A}=A(\hat{z},t,\hbar)\in\mathcal{A}$ with the Weyl symbol $A(z,t,\hbar)$ the following relations hold
\begin{align}
\lim\limits_{\hbar\to 0}\langle \hat{A}\rangle_\Psi=A(Z(t),t,0),
\label{def1a}
\end{align}
\begin{align}
\lim\limits_{\hbar\to 0}\sigma_\Psi(t,\hbar)=\sigma(t).
\label{sig3}
\end{align}
\end{defin}

Here, the functions $Z(t)$ and $\sigma(t)$ are functional parameters of the class ${\mathcal{T}}^t_\hbar(Z(t),\sigma(t))$.

The relation \eqref{def1a} is similar to the definition of the Dirac $\delta$-function in terms of $\delta$-sequence. As is known, $\delta(x-x_0)$ is determined as a linear functional that maps any function $f(x)$ from the specific space (e.g. smooth functions with compact support) to its value at the point $x_0$. In \eqref{def1a}, the analog of the function $f(x)$ is an operator $\hat{A}$ from the set $\mathcal{A}$. The relation \eqref{def1a} maps this operator to its Weyl symbol $A(z,t,\hbar)$ on the trajectory $z=Z(t)$ as $\hbar\to 0$. Thus, the curve $z=Z(t)$ acts as $x_0$, the function $\Psi$ acts as $\delta$-sequence, and the operator $\hat{A}$ along with its Weyl symbol $A(z,t,0)$ acts as a test function. Summing up, for every given moment $t$, the point $z=Z(t)$ is determined in ${\mathbb{R}}^{2n}$, and a function $\Psi\in{\mathcal{T} }^t_\hbar (Z(t),\sigma(t))$ is semiclassically concentrated in a neighborhood of these points that constitute a trajectory $z=Z(t)$ in ${\mathbb{R}}^{2n}$.

In order to obtain the semiclassical evolution of the expectation $\langle\hat{A}(t)\rangle_\Psi$ for a solution $\Psi(\vec{x},t,\hbar)$ to \eqref{hartree1} from the class ${\mathcal{T}}^t_\hbar(Z(t),\sigma(t))$, we go to the limit $\hbar\to 0$ in \eqref{mean2}. In view of \eqref{def1a} and properties of Weyl symbols, we obtain

\begin{align}
&\displaystyle\frac{d }{d t}A(Z(t),t)=-\displaystyle\frac{d\log\sigma(t)}{d t}\cdot
A(Z(t),t)+ \bigg[\displaystyle\frac{\partial A(z,t)}{\partial t} - \big\{ V(z,t),A(z,t)\big\}-\cr
& - 2\Lambda \breve{V}(z,t)A(z,t) +\varkappa \sigma(t) \Big( -\big\{W(z,w,t), A(z,t)\big\}-2\Lambda \breve{W}(z,w,t)A(z,t) \Big)\bigg]\bigg|_{z=w=Z(t)}.
\label{syshet2a}
\end{align}

Here, $\big\{A(z),B(z)\big\}$, $z=(\vec{p},\vec{x})$, stands for Poisson bracket:
\begin{equation}
\big\{A(z),B(z)\big\}=\bigg\langle\displaystyle\frac{\partial A(z)}{\partial z},J
\frac{\partial B(z)}{\partial z}\bigg\rangle=
\sum_{i=1}^n\bigg(\frac{\partial A}{\partial x^i }\frac{\partial B}{\partial p^i }
-\frac{\partial B}{\partial x_i }\frac{\partial A}{\partial p_i }\bigg).
\label{puas1a}
\end{equation}
In \eqref{syshet2a}, we have used the property of the Weyl ordered pseudo-differential operators $A(\hat{z})$, $B(\hat{z})$, $C(\hat{z})$, $D(\hat{z})$ with symbols $A(z)$, $B(z)$, $C(z)$, $D(z)$, respectively, where
\begin{equation}
\begin{gathered}
C(\hat{z})=\big[A(\hat{z}),B(\hat{z}))\big], \qquad D(\hat{z})=\big[A(\hat{z}),B(\hat{z}))\big]_{+},
\end{gathered}
\label{limpo1aa}
\end{equation}
that is as follows \cite{multiind}:
\begin{equation}
\begin{gathered}
\lim_{\hbar\to 0} \dac{C(z)}{i\hbar}=\big\{A(z),B(z)\big\}, \qquad \lim_{\hbar\to 0} D(z)=2A(z)B(z).
\end{gathered}
\label{limpo1a}
\end{equation}



Let us consider two important corollaries of \eqref{syshet2a}.

For the unit operator $\hat{A}=\hat{I}$ with the symbol $A(z)=1$, equation \eqref{syshet2a} yields the following evolution equation for $\sigma(t)$:
\begin{equation}
\dot{\sigma}(t)=-2\Lambda \sigma(t)\big[\breve{V}(Z(t),t)+
\sigma(t)\varkappa\breve{W}(Z(t),Z(t),t)  \big],
\label{sig4}
\end{equation}
where $\dot{\sigma}=d\sigma/dt$.

In view of \eqref{sig4}, for the operator $\hat{A}=\hat{z}$, we obtain the equation for the curve $z=Z(t)$ in ${\mathbb{R}}^{2n}$:
 \begin{equation}
 \dot{Z}(t)=JV_z(Z(t),t)+
\varkappa \sigma(t)JW_z(z,w,t)\mid_{z=w=Z(t)},
\label{syshet5a}
\end{equation}
where $V_z=\partial V(z,t)/\partial z=(\partial V(z,t)/\partial p_i, $ $\partial V(z,t)/\partial x_i)$.

By analogy with \cite{sym2020}, the system \eqref{sig4}, \eqref{syshet5a} is termed the Hamilton--Ehrenfest system with dissipation (HESD) of the first order for the equation \eqref{hartree1} in the class ${\mathcal{T}}^t_\hbar(Z(t),\sigma(t))$.

Note that the system \eqref{sig4}, \eqref{syshet5a} contains the partial information about localization properties of the solutions to \eqref{hartree1} in the space of the dynamical system of the first and zeroth moments of the function $\Psi$. From the derived equations (in particular,  \eqref{sig4}, \eqref{syshet5a}) it is clear that it makes sense to consider only the case $\Lambda=\Or(1)$ within the framework of the semiclassical approximation. If $\Lambda\to \infty$ as $\hbar\to 0$, the solution would rapidly damp so that its dynamics would not be observed at damping times. If $\Lambda\to 0$ as $\hbar\to 0$, we could ignore the non-Hermitian part in the semiclassical approximation that corresponds to the case $\Lambda=0$ in our approach. Note that the last case will be naturally included in our general formalism.

In order to derive explicit analytical expressions for approximate solutions to \eqref{hartree1} in the form of semiclassical asymptotics with respect to the parameter $\hbar$, $\hbar\to 0$, let us consider the class ${\mathcal P}_\hbar^t\big(Z(t),S(t),\sigma(t)\big)$ of functions that singularly depend on asymptotics parameter $\hbar$ \cite{shapovalov:BTS1}:
\begin{align}
{\mathcal{P}}_\hbar^t(Z(t), S(t),\sigma(t))=\bigg\{
\Phi:\Phi(\vec{x},t,\hbar)=\sqrt{\dac{\sigma(t)}{\hbar^{n/2}}}\cdot\varphi\Big(\frac{\Delta\vec{x}}{\sqrt{\hbar}},
t,\hbar\Big) \cdot\exp\Big[\frac{i}{\hbar}\left(S(t)+\langle \vec{P}(t),\Delta \vec{x}  \rangle\right) \Big] \bigg\}.
\label{pth1}
\end{align}
Here, $\Phi(\vec{x},t,\hbar)$ is a general element of the class ${\mathcal{P}}_\hbar^t(Z(t), S(t),\sigma(t))$; the real functions $Z(t)=(\vec{P}(t),\vec{X}(t))$, $S(t)$, and positively defined function $\sigma(t)$ are functional parameters of the class ${\mathcal{P}}_\hbar^t(Z(t), S(t),\sigma(t))$; $\Delta\vec{x}=\vec{x}-\vec{X}(t)$; the function $\varphi(\vec{\xi},t,\hbar)$ belongs to the Schwartz space $\mathbb{S}$ with respect to the variables $\vec{\xi}\in{\mathbb{R}}^n$ and regularly depends on $\hbar$ in a neighborhood of $\hbar=0$; the functions $Z(t), S(t)$, $\varphi(\vec{\xi},t,\hbar)$, and $\sigma(t)$ smoothly depend on $t$. Note that in \cite{shapovalov:BTS1} this class was introduced without normalization factor $\sigma(t)$ since the solutions with conserved norm were considered there. Here, we have added this factor for convenience without loss of generality of the family of classes ${\mathcal P}_\hbar^t\big(Z(t),S(t),\sigma(t)\big)$.

For ease of notation, we will abbreviate ${\mathcal{P}}_\hbar^t(Z(t), S(t),\sigma(t))$ by ${\mathcal{P}}_\hbar^t$ when no confusion will arise.

Functions of the class ${\mathcal{P}}_\hbar^t$ are concentrated at a point moving over the trajectory $z=Z(t)$ in the sense \eqref{def1a}, \eqref{sig3} \cite{shapovalov:BTS1}. They have the special form determined by the ansatz \eqref{pth1}. We will term it as the semiclassical ansatz \eqref{pth1}.

It was proved in \cite{shapovalov:BTS1,sym2020} that, for functions from the class on a finite time interval $t\in[0;T]$, the following asymptotic estimates hold:
\begin{align}
&\{\Delta\hat{z}\}^\alpha=\hat{\Or}(\hbar^{|\alpha|/2}),
\label{estim1a}\\
&\Delta^{(\alpha)}_\Phi(t,\hbar)=\frac{\langle\Phi|\{\Delta\hat{z}\}^\alpha|\Phi\rangle }{\langle\Phi|\Phi\rangle}=\Or(\hbar^{|\alpha|/2}), \quad \Phi\in {\mathcal P}_\hbar^t.
\label{estim1}
\end{align}
Here, the following notations are used. The estimate $\hat{A}=\hat{\Or}(\hbar^{s})$, $s\geq 0$,
in \eqref{estim1a} stands for an operator $\hat{A}$ such that
 \begin{align}
 &\frac{||\hat{A}\Phi||}{||\Phi||}=\Or(\hbar^{s}),\quad \Phi\in {\mathcal{P}}_\hbar^t;
 \label{estim2}
 \end{align}
 $Z(t)=(\vec{P}(t), \vec{X}(t))$, $\Delta\hat{z}=\hat{z}-Z(t)=(\Delta\hat{\vec{p}},\Delta\vec{x})$,
$\Delta\hat{\vec{p}}=\hat{\vec{p}}-\vec{P}(t)$, $\Delta\vec{x}=\vec{x}-\vec{X}(t)$;
$\{\Delta\hat{z}\}^\alpha$ is the operator determined by the Weyl symbol
$\Delta{z}^\alpha$ according to \eqref{pseud1}. The multi-index $\alpha\in {\mathbb{Z}}^{2n}_{+}$ ($2n$-tuple)
reads $\alpha=(\alpha_1,\alpha_2, \ldots , \alpha_{2n})$; $\alpha_j\in {\mathbb{Z}}^{1}_{+}$, $j=1,2,\ldots , 2n$; $|\alpha|=\alpha_1+\alpha+ \ldots  +\alpha_{2n}$. For $v=(v_1,v_2,\ldots, v_{2n})$, $v^{\alpha}=v_1^{\alpha_1}v_2^{\alpha_2}\ldots v_{2n}^{\alpha_{2n}}$. The functions $\Delta^{(\alpha)}_\Phi(t,\hbar)$ are $|\alpha|$-th order central moments of the function $\Phi$.

In particular, we have
 \begin{align}
 &\Delta{x}_j=\hat{\Or}(\sqrt{\hbar}),\quad \Delta\hat{p}_j=\hat{\Or}(\sqrt{\hbar}),
 \quad j=\overline{1,n}, \cr
  &-i\hbar\partial/\partial t -\dot{S}(t)+\langle{\vec{P}}(t),\dot{\vec{X}}(t)\rangle +\langle\dot{Z}(t),J{\Delta\hat{z}} \rangle=\hat{\Or}(\hbar).
  \label{estim3}
  \end{align}
We use the contracted notation $j=\overline{1,n}$ for $j=1,2,\ldots, n$ in \eqref{estim3} and below. Hereinafter, all calculations and commentaries are given for $t\in[0;T]$ where $T<\infty$.

\section{Moments of functions from the class ${\mathcal P}_\hbar^t$}
\label{sec:mom}

In order to construct semiclassical asymptotics for solutions to \eqref{hartree1} in the class \eqref{pth1}, let us introduce the moments of a function $\Psi(\vec{x},t,\hbar)\in {\mathcal{P}}^t_\hbar$ as it follows

\begin{align}
&\sigma_\Psi(t,\hbar)=\langle\Psi|\Psi\rangle(t,\hbar),\quad   Z_\Psi(t,\hbar)=
\frac{1}{\sigma_\Psi(t,\hbar)}\langle\Psi|\hat{z}|\Psi\rangle(t, \hbar)
\label{mom-1}
\end{align}
Here, $\sigma_\Psi(t,\hbar)$ is the zeroth order moment of $\Psi(\vec{x},t,\hbar)$, $Z_\Psi(t,\hbar)$ is the first order moment, and the central moments $\Delta^{(\alpha)}_\Psi(t,\hbar)$ are given by \eqref{estim1}.

Let us present the operators $V(\hat{z},t)$, $\breve{V}(\hat{z},t)$, $W(\hat{z},\hat{w},t)$, and $\breve{W}(\hat{z},\hat{w},t)$  from \eqref{hartree1} in the form of formal power series in $\Delta\hat{z}$ and $\Delta\hat{w}$ in a neighborhood of the trajectory $z=Z(t)$. Then, we have
\begin{equation}
\begin{gathered}
V(\hat{z},t)=V\left(t\right)+\langle V_z\left(t\right), \Delta\hat{z} \rangle + \dac{1}{2}\langle \Delta\hat{z}, V_{zz}\left(t\right) \Delta\hat{z} \rangle+...\\
W(\hat{z},\hat{w},t)=W\left(t\right)+\langle W_z\left(t\right), \Delta\hat{z} \rangle + \langle W_w\left(t\right), \Delta\hat{w} \rangle  + \\
+\dac{1}{2}\langle \Delta\hat{z},  W_{zz}\left(t\right) \Delta\hat{z} \rangle  +\dac{1}{2}\langle \Delta\hat{w},  W_{ww}\left(t\right) \Delta\hat{w} \rangle+\\
+\langle \Delta\hat{z}, W_{zw}\left(t\right) \Delta\hat{w} \rangle+...
\end{gathered}
\label{raz1}
\end{equation}
Hereinafter, the following notations are used:
\begin{equation}
\begin{gathered}
V(t)=V\left(Z(t),t\right), \qquad V_z\left(t\right)=\left(\dac{\pa V\left(z,t\right)}{\pa z_k}\Big|_{z=Z(t)}\right), \qquad V_{zz}\left(t\right)=\left(\dac{ \pa^2 V\left(z,t\right)}{\pa z_k \pa z_j}\Big|_{z=Z(t)}\right), \\
W(t)=W\left(Z(t),Z(t),t\right), \qquad W_z\left(t\right)=\left(\dac{\pa W\left(z,w,t\right)}{\pa z_k}\Big|_{z=w=Z(t)}\right), \qquad k,j=\overline{1,2n},
\end{gathered}
\label{raz2}
\end{equation}
and so on by analogy. The expansions for $\breve{V}(\hat{z},t)$ and $\breve{W}(\hat{z},\hat{w},t)$ are similar to \eqref{raz1}. In view of \eqref{estim1a}, the part of the operators dropped \eqref{raz2} is estimated as $\hat{{\rm O}}(\hbar^{(m+1)/2})$ where $m$ is the greatest power of a polynomial of $\Delta \hat{z}$, $\Delta \hat{w}$ that was taken into consideration in the asymptotic expansion. In particular, if we limit ourselves to terms that were explicitly written in \eqref{raz1}, then $m=2$.

Let us denote the aggregate vector of moments of the function $\Psi(\vec{x},t,\hbar)$ by
\begin{equation}
g_{\Psi}(t,\hbar)=\left(\sigma_{\Psi}(t,\hbar),Z_{\Psi}(t,\hbar),\Delta_{\Psi}^{(\alpha)}(t,\hbar)\right), \qquad |\alpha|=\overline{2,\infty}.
\label{momg1}
\end{equation}
It is known that $g_{\Psi}(t,\hbar)$ determines the function $\Psi(\vec{x},t,\hbar)$ (see, e.g., \cite{bojowald05}). However, the system of equations for \eqref{momg1} is an infinite set of ODEs.

Let us also introduce the aggregate vector $g_{\Psi}^{(M)}(t,\hbar)$ that reads
\begin{equation}
g_{\Psi}^{(M)}(t,\hbar)=\left(\sigma_{\Psi}(t,\hbar),Z_{\Psi}(t,\hbar),\Delta_{\Psi}^{(\alpha)}(t,\hbar)\right)+{\rm O}(\hbar^{(M+1)/2}), \qquad |\alpha|=\overline{2,M}.
\label{momg2}
\end{equation}
In \eqref{momg2}, we suppose that all moments included in $g_{\Psi}^{(M)}(t,\hbar)$ are determined within accuracy of ${\rm O}(\hbar^{(M+1)/2})$ where $M$ indicates both the greatest order of moments included in $g_{\Psi}^{(M)}(t,\hbar)$ and their accuracy in this aggregate vector. Substituting \eqref{raz1} in \eqref{sig2}, \eqref{mean2} and using estimates \eqref{estim1a}, \eqref{estim1}, one gets the closed system of ODEs of the form
\begin{equation}
\dot{g}_{\Psi}^{(M)}(t,\hbar)=\Gamma^{(M)}\left(g_{\Psi}^{(M)}(t,\hbar),t,\hbar\right).
\label{momg3}
\end{equation}
Finally, let us introduce the aggregate vector $g^{(M)}(t,\hbar,{\bf C})$ that is a particular solution of
\begin{equation}
\dot{g}^{(M)}(t,\hbar,\BFC)=\Gamma^{(M)}\left(g^{(M)}(t,\hbar,\BFC),t,\hbar\right).
\label{momg4}
\end{equation}
with integration constant $\BFC$. The system \eqref{momg4} is termed the HESD of the $M$-th order.

Note that, in view of properties of the class ${\mathcal{P}}_\hbar^t$, the following relations hold \cite{shapovalov:BTS1}:
\begin{equation}
\begin{gathered}
Z_{\Psi}(t,\hbar)=Z^{(0)}_{\Psi}(t)+\hbar Z_{\Psi}^{(1)}(t)+{\rm O}(\hbar^{3/2}), \\
\sigma_{\Psi}(t,\hbar)=\sigma_{\Psi}^{(0)}(t)+\hbar\sigma_{\Psi}^{(1)}(t)+{\rm O}(\hbar^{3/2}).
\end{gathered}
\label{svmom1}
\end{equation}

From \eqref{def1a}, \eqref{sig3}, it also follows that
\begin{equation}
Z^{(0)}_{\Psi}(t)=Z(t), \qquad \sigma^{(0)}_{\Psi}(t)=\sigma(t).
\label{svmom2}
\end{equation}
The asymptotic solutions can be obtained in those classes ${\mathcal P}_\hbar^t\big(Z(t),S(t),\sigma(t)\big)$ that are defined by functions determined by equations \eqref{sig4}, \eqref{syshet5a} and a function $S(t)$ given by
\begin{equation}
\dot{S}(t)=\langle \vec{P}(t),\dot{\vec{X}}(t)\rangle - V(t)-\varkappa\sigma(t)W(t).
\label{deist1}
\end{equation}

Due to the uniqueness of the solution to the Cauchy problem for \eqref{momg3} and \eqref{momg4}, we have $g_{\Psi}^{(M)}(t,\hbar)=\dot{g}^{(M)}(t,\hbar,\BFC)$ under the algebraic condition
\begin{equation}
\BFC=\BFC[\Psi].
\label{momalg1}
\end{equation}

In view of \eqref{svmom2}, the algebraic condition \eqref{momalg1} is degenerate for the integration constants of the HESD of the first order \eqref{sig4}, \eqref{syshet5a} in the class ${\mathcal{P}}_\hbar^t$ with the given $Z(t)$ and $\sigma(t)$. Hence, the functions $Z(t)=Z^{(0)}(t)$ and $\sigma(t)=\sigma^{(0)}(t)$ do not depend on $\BFC$.

Thus, equations \eqref{sig4}, \eqref{syshet5a} form the system \eqref{momg4} for $M=1$. Next, we will show that we have to solve at least the HESD of the second order to construct the leading term of asymptotics for the solution $\Psi(\vec{x},t)$ to \eqref{hartree1}. In order to construct higher order asymptotics, we must solve the HESD of a higher order. The equations of the HESD of the second order are given explicitly in Appendix \ref{sec:app1}.

\section{Associated linear Schr\"odinger equation with dissipation}
\label{sec:alse}

In the class ${\mathcal{P}}_\hbar^t\left(Z(t),S(t),\sigma(t)\right)$, the solution to \eqref{hartree1} can be sought in the form
\begin{equation}
\Psi(\vec{x},t,\hbar)=\sum_{k=0}^{M}\hbar^{k/2}\Psi^{(k)}(\vec{x},t,\hbar)+\bar{\Or}(\hbar^{(M+1)/2}),
\label{aseq1}
\end{equation}
where $\Psi^{(k)}(\vec{x},t,\hbar)\in {\mathcal{P}}_\hbar^t\left(Z(t),S(t),\sigma(t)\right)$, $k=\overline{0,M}$, and the estimates $f(\vec{x},t,\hbar)=\bar{\Or}(\hbar^s)$, $s\geq 0$, means
\begin{equation}
\max_{t\in [0;T]}||f(\vec{x},t,\hbar)||=\Or(\hbar^s).
\label{ocnorm1}
\end{equation}

Let us write \eqref{hartree1} as follows:
\begin{equation}
\hat{L}[\Psi]\Psi(\vec{x},t,\hbar)=\bigg\{ -i\hbar\pa_t + H(\hat{z},t)[\Psi]-i\hbar \Lambda \breve{H}(\hat{z},t)[\Psi] \bigg\}\Psi(\vec{x},t)=0.
\label{aseq2}
\end{equation}
Following \cite{shapovalov:BTS1}, we expand the operators $H(\hat{z},t)[\Psi]$ and $\breve{H}(\hat{z},t)[\Psi]$ in a series in $\Delta \hat{z}$, expand the kernels of nonlinear terms in a series in $\Delta\hat{w}$ (see \eqref{raz1}), and replace the respective integrals to the moments of the function $\Psi$. Then, in view of \eqref{estim1a}, \eqref{estim1}, we obtain the expansion of the operator $\hat{L}[\Psi]$ in the form
\begin{equation}
\hat{L}[\Psi]=\sum_{k=0}^{M}\hat{L}^{(k)}(g_\Psi(t,\hbar))+\hat{\Or}(\hbar^{(M+1)/2}),
\label{opraz1}
\end{equation}
where $\hat{L}^{(k)}(g_\Psi(t,\hbar))=\hat{\Or}(\hbar^{k/2})$, $\hbar\to 0$, $k=\overline{0,M}$.

Using \eqref{aseq2}, \eqref{opraz1}, we can write \eqref{aseq2} as follows:
\begin{equation}
\begin{gathered}
\bigg(\sum_{k=0}^{M}\hat{L}^{(k)}(g_\Psi(t,\hbar)) \bigg)\bigg(\sum_{m=0}^{M}\hbar^{m/2}\Psi^{(m)}(\vec{x},t,\hbar)\bigg)=\bar{\Or}(\hbar^{(M+1)/2}).
\end{gathered}
\label{aseq3}
\end{equation}

Since the operators $\hat{L}^{(k)}(g_\Psi(t,\hbar))$ smoothly depend on the argument $g_\Psi(t,\hbar)$, in view of \eqref{momg2}, the operators $\hat{L}^{(k)}(g_\Psi(t,\hbar))$ can be substituted for $\hat{L}^{(k)}(g_\Psi^{(M)}(t,\hbar))$ in \eqref{aseq3}.

It can be shown that $\hat{L}[\Psi]=\hat{\Or}(\hbar)$ in the class ${\mathcal{P}}_\hbar^t$, i.e. the expansion \eqref{opraz1} can be chosen so that
\begin{equation}
\begin{gathered}
\hat{L}^{(0)}(g_\Psi(t,\hbar))=0,\\
\hat{L}^{(1)}(g_\Psi(t,\hbar))=0.
\end{gathered}
\label{ekviv1}
\end{equation}
The first identity in \eqref{ekviv1} follows from \eqref{estim3} and \eqref{deist1}, while the second one follows from \eqref{estim3} and \eqref{syshet5a}.

Thus, \eqref{aseq3} can be written as follows:
\begin{equation}
\begin{gathered}
\bigg(\sum_{k=0}^{M-2}\hat{L}^{(k+2)}(g_\Psi^{(M)}(t,\hbar)) \bigg) \bigg(\sum_{m=0}^{M}\hbar^{m/2}\Psi^{(m)}(\vec{x},t,\hbar)\bigg)=\bar{\Or}(\hbar^{(M+1)/2}).
\end{gathered}
\label{raznov1}
\end{equation}

Grouping terms of various orders with respect to $\hbar$ in \eqref{raznov1}, we derive the system of equations for functions $\Psi^{(k)}$, $k=\overline{0,M-2}$:
\begin{equation}
\begin{array}{l}
\hbar^1:\quad \hat{L}^{(2)}(g_\Psi^{(M)}(t,\hbar)) \Psi^{(0)}(\vec{x},t,\hbar)=0,\\
\hbar^{3/2}:\quad \hbar^{1/2}\hat{L}^{(2)}(g_\Psi^{(M)}(t,\hbar)) \Psi^{(1)}(\vec{x},t,\hbar)+\hat{L}^{(3)}(g_\Psi^{(M)}(t,\hbar)) \Psi^{(0)}(\vec{x},t,\hbar)=0,\\
...\\
\hbar^{M/2}:\quad \hbar^{(M-2)/2}\hat{L}^{(2)}(g_\Psi^{(M)}(t,\hbar)) \Psi^{(M-2)}(\vec{x},t,\hbar)+\\
+\hbar^{(M-3)/2}\hat{L}^{(3)}(g_\Psi^{(M)}(t,\hbar)) \Psi^{(M-1)}(\vec{x},t,\hbar)+...+\hat{L}^{(M)}(g_\Psi^{(M)}(t,\hbar)) \Psi^{(0)}(\vec{x},t,\hbar)=0.
\end{array}
\label{raznov2}
\end{equation}
In particular, for $M=2$, one readily gets that \eqref{raznov2} yields the following single equation for the leading term of asymptotics:
\begin{equation}
\begin{gathered}
\hat{L}^{(2)}(g_\Psi^{(2)}(t,\hbar)) \Psi^{(0)}(\vec{x},t,\hbar)=0.
\end{gathered}
\label{aseq4}
\end{equation}
Following \cite{shapovalov:BTS1}, we term $\Psi^{(0)}(\vec{x},t,\hbar)$ by the solution with right-hand side accuracy of $\Or(\hbar^{3/2})$ in the sense that it is generated by \eqref{raznov1} with accuracy of $\bar{\Or}(\hbar^{3/2})$. Hereinafter, we suppose that $\hat{L}_2$ is the operator from the expansion \eqref{opraz1} satisfying \eqref{ekviv1}.

Let us introduce the auxiliary linear equation
\begin{equation}
\hat{L}^{(2)}(g^{(M)}(t,\hbar,\BFC))\Phi(\vec{x},t,\hbar,\BFC)=0, \qquad M\geq 2,
\label{aseq6}
\end{equation}
with a solution $\Phi(\vec{x},t,\hbar,\BFC)\in {\mathcal{P}}_\hbar^t\left(Z(t),S(t),\sigma(t)\right)$.

Then, the following proposition holds \cite{shapovalov:BTS1}
\begin{demo}
\label{teo1}
Let $\Psi(\vec{x},t,\hbar)\in {\mathcal{P}}_\hbar^t\left(Z(t),S(t),\sigma(t)\right)$ be a solution to the Cauchy problem for \eqref{hartree1} with the initial condition
\begin{equation}
\Psi(\vec{x},t,\hbar)\Big|_{t=0}=\varphi(\vec{x},\hbar),
\label{demo1}
\end{equation}
and $\Phi(\vec{x},t,\hbar,\BFC[\varphi])$ be a solution to the Cauchy problem for \eqref{aseq6}, $\BFC=\BFC[\varphi]$, with the initial condition
\begin{equation}
\Phi(\vec{x},t,\hbar,\BFC[\varphi])\Big|_{t=0}=\varphi(\vec{x},\hbar).
\label{demo2}
\end{equation}
Then, we have
\begin{equation}
\Phi(\vec{x},t,\hbar,\BFC[\varphi])-\Psi(\vec{x},t,\hbar)=\bar{\Or}(\sqrt{\hbar}).
\end{equation}
\end{demo}

This proposition follows from the uniqueness of the solution to the Cauchy problems for \eqref{momg4} and \eqref{aseq6}.

To put it differently, the leading term of the asymptotic solution to the Cauchy problem for the original nonlinear equation \eqref{hartree1} in the class ${\mathcal{P}}_\hbar^t$ can be found among solutions to $\BFC$-parametric family of linear equations \eqref{aseq6}. The search of the appropriate solution is reduced to the algebraic conditions
\begin{equation}
\BFC=\BFC[\varphi].
\label{demo3}
\end{equation}
The parameters $\BFC$ (integration constants for \eqref{momg4}) that meet the condition \eqref{demo3} can be determined by initial conditions for the moments of the function $\Psi$, $g_{\Psi}^{(M)}(t,\hbar)\big|_{t=0}$.

Following \cite{sym2020}, we term \eqref{aseq6} as the associated linear Schr\"{o}dinger equation with dissipation (ALSED). It can be written as follows:
\begin{equation}
\begin{gathered}
\hat{L}^{(2)}(g^{(M)}(t,\hbar,\BFC))\Phi(\vec{x},t,\hbar)=\\
=\bigg\{-i\hbar\pa_t + H(t,\BFC,\hbar) + \langle H_z(t),\Delta \hat{z} \rangle + \dac{1}{2} \langle \Delta \hat{z}, H_{zz}(t) \Delta \hat{z} \rangle \bigg\} \Phi(\vec{x},t,\hbar)=0,
\end{gathered}
\label{alsd1}
\end{equation}
where
\begin{equation}
\begin{array}{l}
H(t,\BFC,\hbar)=H^{(0)}(t)+\hbar H^{(1)}(t,\BFC), \cr \cr
H^{(0)}(t)=V(t)+\varkappa \sigma (t) W(t), \cr \cr
H^{(1)}(t,\BFC)=\dac{\varkappa}{2}\sigma(t) \Sp[W_{ww}(t)\cdot \Delta_2^{(1)}(t,\BFC)]+\varkappa \sigma(t) \langle W_w(t), Z^{(1)}(t,\BFC) \rangle +\cr
+ \varkappa \sigma^{(1)}(t,\BFC) W(t) -i\Lambda \breve{V}(t)-i\Lambda \varkappa \sigma(t) \breve{W}(t), \cr \cr
H_z(t)=V_z(t)+\varkappa \sigma(t) W_{z}(t), \cr \cr
H_{zz}(t)=V_{zz}(t)+\varkappa \sigma(t) W_{zz}(t).
\end{array}
\label{alsd2}
\end{equation}
Here, $\Delta_{2,ij}^{(1)}(t)=\dac{1}{2\hbar}\langle \Delta \hat{z}_i \Delta \hat{z}_j+\Delta \hat{z}_j \Delta \hat{z}_i\rangle$ is the dispersion matrix (matrix of the second order central moments).

Green's function for \eqref{alsd1}, \eqref{alsd2} reads \cite{bagrov1}

\begin{equation}
\begin{gathered}
G(\vec{x},\vec{y},t,{\bf C},\hbar)=\displaystyle\frac{1}{\sqrt{\det\big(-2\pi i \hbar M_3(t)\big)}} \exp\Bigg\{\displaystyle\frac{i}{\hbar}\bigg[\displaystyle\int\limits_0^t \Big( \langle \vec{P}(\tau),\dot{\vec{X}}(\tau) \rangle-H(\tau,{\bf C},\hbar) \Big)d\tau+ \\
+\langle \vec{P}(t), \Delta \vec{x} \rangle - \langle \vec{P}(0), \Delta \vec{y} \rangle - \displaystyle\frac{1}{2} \langle \Delta \vec{x}, M_3^{-1}(t) M_1(t)\Delta \vec{x} \rangle +\\
+\langle \Delta \vec{x}, M_3^{-1}(t) \Delta \vec{y} \rangle - \displaystyle\frac{1}{2} \langle \Delta \vec{y}, M_4(t)M_3^{-1}(t)\Delta \vec{y} \rangle \bigg] \Bigg\}, \label{ur7}
\end{gathered}
\end{equation}
where $\Delta y = y-\vec{X}(0)$, and $2n\times 2n$ matrix $M(t)=\begin{pmatrix} M_1(t) && -M_3(t) \cr -M_2(t) && M_4(t) \end{pmatrix}$ is a solution to the Cauchy problem
\begin{equation}
\dot{M}=-M\cdot H_{zz}\big(t\big)J, \qquad M(0)=\begin{pmatrix} I_{n\times n} && 0 \cr 0 && I_{n\times n} \end{pmatrix}.
\label{ur5}
\end{equation}
This function generates the following semiclassical evolution operator for asymptotic solutions from the class ${\mathcal{P}}_\hbar^t\left(Z(t),S(t),\sigma(t)\right)$:
\begin{equation}
\Psi^{(0)}(\vec{x},t)=\hat{U}(t)\varphi(\vec{x})=\dil_{{\mathbb{R}}^n}G(\vec{x},\vec{y},t,{\bf C}[\varphi]) \varphi(\vec{y})d\vec{y}, \qquad \Psi^{(0)}(\vec{x},0)=\varphi(\vec{x}).
\label{green1}
\end{equation}
Note that the semiclassical evolution operator $\hat{U}(t)$ is nonlinear since the integrand in \eqref{green1} depends nonlinearly on $\varphi(\vec{x})$ by way of $\BFC[\varphi]$.

\section{Semiclassical symmetry operators}
\label{sec:sym}

The solutions to the equation \eqref{alsd1}, \eqref{alsd2} that determine the asymptotic solutions to \eqref{hartree1} can be generated using symmetry operators.

Let $a(t)\in{\mathbb{C}}^{2n}$ be a solution to
\begin{equation}
\dot{a}=J H_{zz}(t) a.
\label{opsym1}
\end{equation}
Then, the operator
\begin{equation}
\hat{a}(t)=\langle a(t), \Delta \hat{z}\rangle
\label{opsym2}
\end{equation}
is a symmetry operator of the first order for the equation \eqref{alsd1}, \eqref{alsd2}.

In view of the explicit form for the evolution operator \eqref{ur7}, \eqref{green1}, an asymptotic solution to \eqref{hartree1} in the class ${\mathcal{P}}_\hbar^t\left(Z(t),S(t),\sigma(t)\right)$ can be written as follows:
\begin{equation}
\Psi^{(0)}(\vec{x},t)=\exp\bigg[-\dac{i}{\hbar}\dil_{0}^{t}H(\tau,\BFC[\psi])d\tau\bigg]\psi(\vec{x},t).
\label{opsym2}
\end{equation}

Then, the function $\tilde{\Psi}^{(0)}(\vec{x},t)$ given by
\begin{equation}
\begin{gathered}
\tilde{\Psi}^{(0)}(\vec{x},t)=\exp\bigg[-\dac{i}{\hbar}\dil_{0}^{t}H(\tau,\BFC[\tilde{\psi}])d\tau\bigg]\tilde{\psi}(\vec{x},t),\\
\tilde{\psi}(\vec{x},t)=\hat{a}(t)\psi(\vec{x},t),
\end{gathered}
\label{opsym3}
\end{equation}
is also an asymptotic solution to \eqref{hartree1}. In terms of the evolution operator $\hat{U}(t)$ \eqref{green1}, the relation \eqref{opsym3} reads

\begin{equation}
\tilde{\Psi}^{(0)}(\vec{x},t)=\hat{U}(t)\hat{a}(t)\hat{U}^{-1}(t)\Psi^{(0)}(\vec{x},t).
\label{opsym4}
\end{equation}
The relation \eqref{opsym4} indicates clearly that the functions $\tilde{\Psi}^{(0)}$ and $\Psi^{(0)}$ are related nonlinearly due to the nonlinearity of $\hat{U}(t)$.

Let $a_k(t)\in{\mathbb{C}}^{2n}$, $k\in {\mathbb{N}}$, be linearly independent solutions to \eqref{opsym2} satisfying the skew-orthogonality condition
\begin{equation}
\{a_k,a_m\}=\{a_k^*,a_m^*\}=0, \qquad \{a_k^*,a_m\}=2i\delta_{km}, \qquad \forall k,m\in{\mathbb{N}},
\label{opsym5}
\end{equation}
where $\{a_1,a_2\}=\langle a_1, J a_2 \rangle$, and $\delta_{km}$ the Kronecker delta. Then, the respective symmetry operators for the ALSE
\begin{equation}
\hat{a}_k(t)=\dac{1}{\sqrt{2\hbar}}\langle a_k(t), J \Delta \hat{z}\rangle, \qquad \hat{a}_k^{+}(t)=\dac{1}{\sqrt{2\hbar}}\langle a_k^*(t), J \Delta \hat{z}\rangle,
\label{opsym6}
\end{equation}
form Heisenberg's Lie algebra:
\begin{equation}
[\hat{a}_k,\hat{a}_m]=[\hat{a}_k^+,\hat{a}_m^+]=0, \qquad [\hat{a}_k,\hat{a}_m^{+}]=\delta_{km}, \qquad \forall k,m\in{\mathbb{N}}.
\label{opsym7}
\end{equation}

Thus, the linear symmetry operators \eqref{opsym6} for ALSE that form the Lie algebra \eqref{opsym7} generate the nonlinear approximate symmetry operators for the original nonlinear equation \eqref{hartree1}. Using the set of operators \eqref{opsym6}, one can construct analogs of the well-known Fock states \cite{schweber61} for the nonlinear equation \eqref{hartree1}.

\section{Example}
\label{sec:example}

Hereunder, we illustrate the formalism proposed with the simple but nontrivial example. Let us consider the model equation
\begin{equation}
\begin{gathered}
\bigg\{-i\hbar\pa_t + c_1\hat{p}^2  + c_2 \varkappa\dil_{{\mathbb{R}}} \exp\left(-\dac{(x-y)^2}{\gamma^2}\right)|\Psi(\vec{y},t)|^2dy-\\
-i\hbar\Lambda\bigg[-\epsilon+\hat{p}^2  +\varkappa \dil_{{\mathbb{R}}} \exp\left(-\dac{(x-y)^2}{\gamma^2}\right)|\Psi(\vec{y},t)|^2dy\bigg]\bigg\}\Psi(\vec{x},t)=0.
\end{gathered}
\label{prim1}
\end{equation}
It was derived in \cite{arecchi2000} for description of the field of the BEC in an atom laser that is a fundamentally open system. This equation is the reduction of the system of two related equations. The first one, the GPE, describes the field $\Psi$ of condensed atoms. The second one, the reaction-diffusion equation, describes the density of uncondensed atoms. As we noted earlier, we operate with a nonlocal form of the nonlinearity within the framework of our formalism. The cases of $c_1>0$, $\epsilon>0$, $\Lambda>0$, and $c_2\gtrless 0$ were considered in \cite{arecchi2000}. These coefficients depend on the coupling constant between condensed and uncondensed atoms \cite{kneer98}, parameters of the laser pumping, parameters of a trap, and properties of atoms themselves such as the effective mass and self interaction strength.

In our notations, we have
\begin{equation}
\begin{gathered}
V(z,t)=c_1 p^2,\\
W(z,w,t)=c_2 \exp\left(-\dac{(x-y)^2}{\gamma^2}\right),\\
\breve{V}(z,t)=-\epsilon+p^2,\\
\breve{W}(z,w,t)=\exp\left(-\dac{(x-y)^2}{\gamma^2}\right),\\
z=(p,x),\qquad w=(p_y,y).
\end{gathered}
\label{prim2}
\end{equation}

Let us pose the initial condition of the form
\begin{equation}
\Psi(x,t)\Big|_{t=0}=\varphi(x)=\sqrt{\dac{N}{\zeta \sqrt{\pi \hbar}}}\exp\left(-\dac{x^2}{2\hbar \zeta^2}\right),
\label{prim3}
\end{equation}
that implies that some amount of condensate is trapped at the initial moment of time.

From \eqref{prim3}, in view of \eqref{mom-1}, \eqref{app1}, \eqref{demo3}, the initial conditions for HESD \eqref{sig4}, \eqref{syshet5a}, \eqref{sge1}, \eqref{sge2}, \eqref{sge4} read
\begin{equation}
\begin{gathered}
\sigma(0)=\sigma^{(0)}(0)=N, \qquad \sigma^{(1)}(0)=0, \qquad Z(0)= Z^{(0)}(0)=0, \qquad Z^{(1)}(0)=0,\\
\Delta^{(1)}_2(0)=\begin{pmatrix} \dac{1}{2\zeta^2} & 0 \\ 0 & \dac{\zeta^2}{2} \end{pmatrix}.
\end{gathered}
\label{prim4}
\end{equation}
In this section, we will omit the argument $\BFC$ since the initial conditions \eqref{prim4}, which are integration constants $\BFC$, will be explicitly included in expressions.

In view of \eqref{prim4} and symmetries in coefficients \eqref{prim2}, one readily gets
\begin{equation}
Z(t)=0, \qquad Z^{(1)}(t)=0.
\label{prim5}
\end{equation}

The Cauchy problem for \eqref{sig4} is given by
\begin{equation}
\dot{\sigma}(t)=-2\Lambda \sigma(t) \Big(-\epsilon+ \varkappa \sigma(t)\Big), \qquad \sigma(0)=N,
\label{prim6}
\end{equation}
and its solutions reads
\begin{equation}
\sigma(t)=\dac{N \epsilon e^{2\Lambda \epsilon t}}{\epsilon+N\varkappa \left(e^{2\Lambda \epsilon t}-1\right)}.
\label{prim7}
\end{equation}
Note that, in our formalism, the function $\sigma(t)$ must be positively defined. Hence, for $\Lambda>0$, $\epsilon<0$, and $N\varkappa<\epsilon$, the asymptotics can be constructed only for $0\leq t < \dac{1}{2\Lambda \epsilon}\ln\left(1-\dac{\epsilon}{N\varkappa}\right)$. The consideration of a finite time interval when constructing asymptotics is reasonable since the exact solutions to the original nonlinear equation do not necessarily exist for infinite time interval.

The condition $\epsilon>0$ means that the threshold condition (gain=losses) is met \cite{arecchi2000}. The formula \eqref{prim7} shows that, for great $t$, in zeroth approximation by $\hbar$, the evolution of the condensate in a trap is affected by the effective pump (effective implies that $\epsilon$ takes account of both the gain and losses) and nonlinearity factor. The positive nonlinearity factor, $\varkappa>0$, corresponds to the respective interatomic interaction. The increase in the effecting pump leads to the increase in the amount of condensate, while the increase in the respectively interaction strength leads to the decrease in the amount of condensate.

Next, we calculate the matrix coefficient
\begin{equation}
H_{zz}(t)=\begin{pmatrix}2c^1 & 0 \\ 0 & -\dac{2c_2\varkappa \sigma(t)}{\gamma^2}\end{pmatrix}.
\label{prim8}
\end{equation}
In view of the symmetry of the matrix $\Delta_2^{(1)}(t)$, we denote
\begin{equation}
\Delta_2^{(1)}(t)=\begin{pmatrix}\alpha_{pp}(t) & \alpha_{px}(t) \\ \alpha_{px}(t) &  \alpha_{xx}(t)\end{pmatrix}.
\label{prim9}
\end{equation}

Then, equation \eqref{sge1} reads
\begin{equation}
\begin{gathered}
\dac{d}{dt}\begin{pmatrix} \alpha_{pp}(t) \\ \alpha_{px}(t) \\ \alpha_{xx}(t) \end{pmatrix}=
\begin{pmatrix} 0 & 4 c_2 \varkappa \gamma^{-2} \sigma(t) & 0 \\
2c_1 & 0 & 2 c_2 \varkappa \gamma^{-2}\sigma(t) \\
0 & 4c_1 & 0 \end{pmatrix}
\begin{pmatrix}  \alpha_{pp}(t) \\ \alpha_{px}(t) \\ \alpha_{xx}(t) \end{pmatrix},\\
\begin{pmatrix}  \alpha_{pp}(0) \\ \alpha_{px}(0) \\ \alpha_{xx}(0) \end{pmatrix}=\dac{1}{2}\begin{pmatrix}  \zeta^{-2} \\ 0 \\ \zeta^{2} \end{pmatrix},
\end{gathered}
\label{prim10}
\end{equation}
while equation \eqref{sge4} reads
\begin{equation}
\begin{gathered}
\dot{\sigma^{(1)}}(t)=-2\Lambda\sigma(t)\alpha_{pp}(t)c_1-2\Lambda \varkappa \sigma(t) \sigma^{(1)}(t) (1+ c_2),\\
\sigma^{(1)}(0)=0.
\end{gathered}
\label{prim11}
\end{equation}

The solution to \eqref{prim11} is given by
\begin{equation}
\begin{gathered}
\sigma^{(1)}(t)=-2 \Lambda c_1 v^{-1}(t)\dil_{0}^{t} v(\tau)  \sigma(\tau) \alpha_{pp}(\tau) d\tau,\\
v(t)=\left(\epsilon+N\varkappa\left(e^{2\Lambda \epsilon t}-1\right)\right)^{1+c_2},
\end{gathered}
\label{prim12}
\end{equation}

The system\eqref{ur5} can be written as
\begin{equation}
\dot{M}(t)=M(t) \begin{pmatrix} 0 & 2c^1 \\ \dac{2 c_2 \varkappa \sigma(t)}{\gamma^2} & 0\end{pmatrix}, \qquad M(0)=\begin{pmatrix}1 & 0 \\ 0 & 1 \end{pmatrix}.
\label{prim13}
\end{equation}

Note that the solutions to equation \eqref{sge1} (that takes the form of \eqref{prim9}, \eqref{prim10} in this particular example) can be expressed via the solutions to \eqref{ur5} (that is given by \eqref{prim13} in this example) as follows:
\begin{equation}
\Delta_2^{(1)}(t)=M^{\top}(t)\Delta_2^{(1)}(0)M(t).
\label{prim14}
\end{equation}

The exact solutions to \eqref{prim13} can be expressed via the Meijer G-function. It is quite cumbersome and, for that reason, is given in Appendix \ref{app1}.

Green's function \eqref{ur7} reads
\begin{equation}
\begin{gathered}
G(x,y,t)=\displaystyle\frac{1}{\sqrt{-2\pi i \hbar M_3(t)}} \exp\Bigg\{\displaystyle-\frac{i}{\hbar}\bigg[\displaystyle\int\limits_0^t H(\tau)d\tau+ \\
 +\dac{1}{2M_3(t)}\left( M_1(t) x^2  -2xy + M_4(t) y^2\right) \bigg] \Bigg\},\\
 H(t)=c_2 \varkappa \sigma(t) - \hbar c_2 \varkappa\sigma(t)\dac{\alpha_{xx}(t)}{\gamma^2} +\hbar c_2 \varkappa \sigma^{(1)}(t) + i\hbar\Lambda \epsilon - i\hbar\Lambda\varkappa \sigma(t).
\end{gathered}
\label{prim15}
\end{equation}

The substitution of \eqref{prim15} and \eqref{prim3} into \eqref{green1} yields
\begin{equation}
\begin{gathered}
\Psi^{(0)}(x,t)=\sqrt{\dac{N \zeta}{ \Big(M_4(t)\zeta^2 -i M_3(t)\Big)\sqrt{\pi \hbar}}} \exp\Bigg\{\displaystyle-\frac{i}{\hbar}\displaystyle\int\limits_0^t H(\tau)d\tau\Bigg\}\times \\
\times \exp\Bigg\{-\dac{x^2}{2\hbar}\cdot\dac{\zeta^2\Big(1-M_1(t)M_4(t)\Big)+iM_1(t)M_3(t)}{M_3(t)\Big(M_3(t)+iM_4(t) \zeta^2\Big)} \Bigg\}.
\end{gathered}
\label{prim16}
\end{equation}

\begin{figure}[h]
\centering\begin{minipage}[b][][b]{0.6\linewidth}\centering
    \includegraphics[width=10.5 cm]{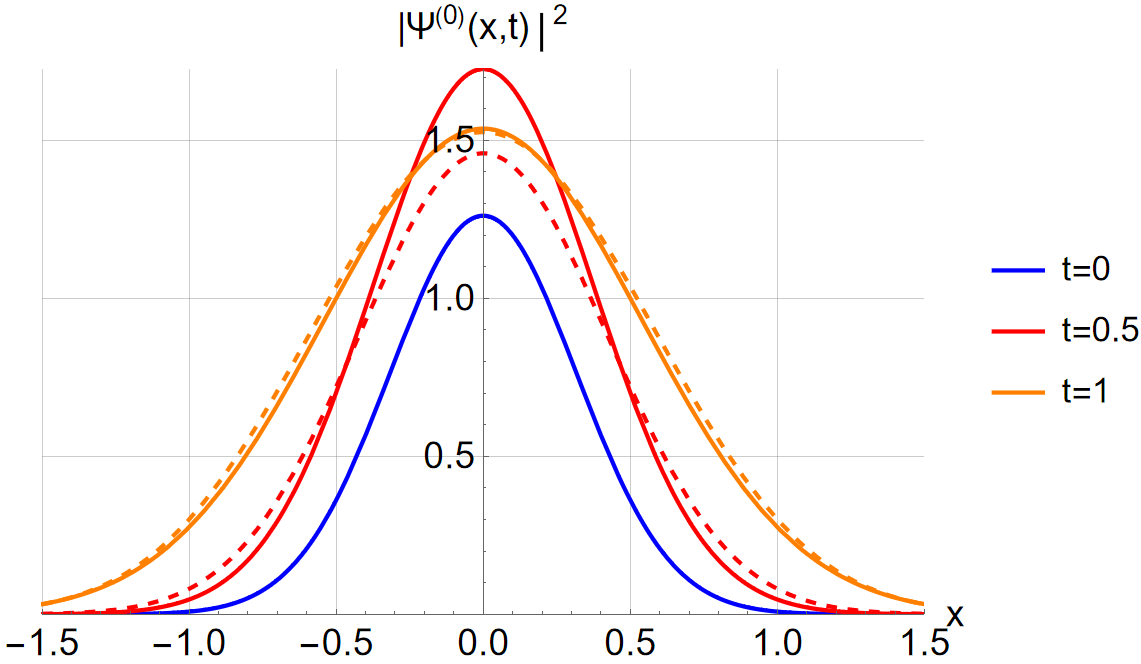} \\ a) $\Lambda=2$, $\varkappa=0.2$
  \end{minipage}\\
 \begin{minipage}[b][][b]{0.6\linewidth} \centering
    \includegraphics[width=10.5 cm]{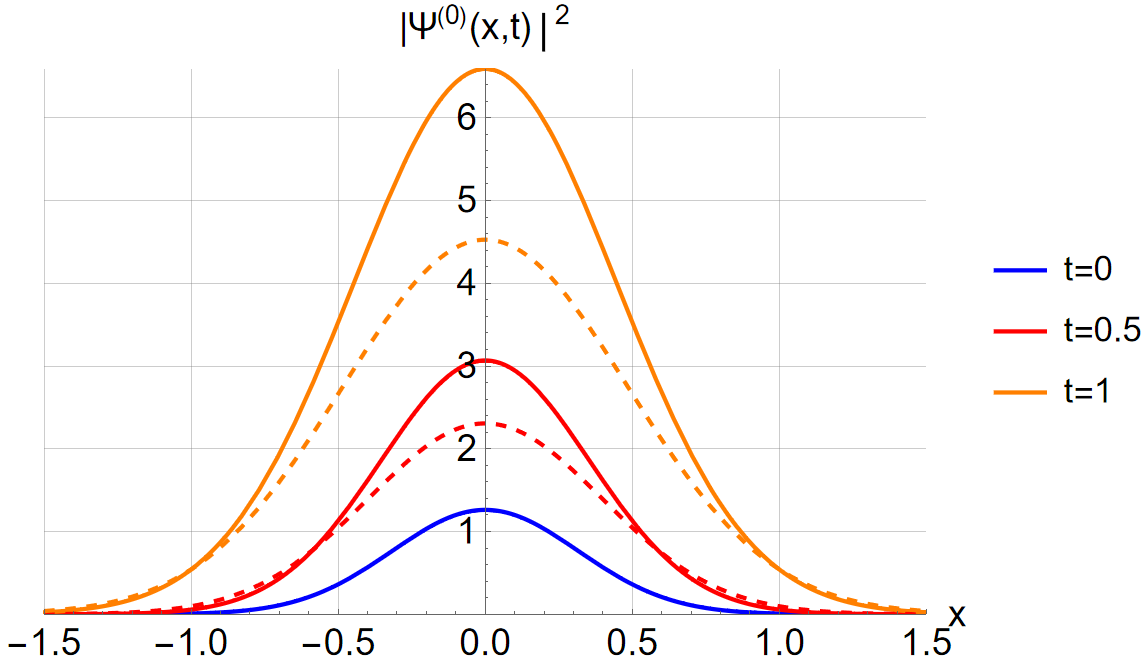} \\ b) $\Lambda=2$, $\varkappa=0$
  \end{minipage}\\
\begin{minipage}[b][][b]{0.6\linewidth} \centering
    \includegraphics[width=10.5 cm]{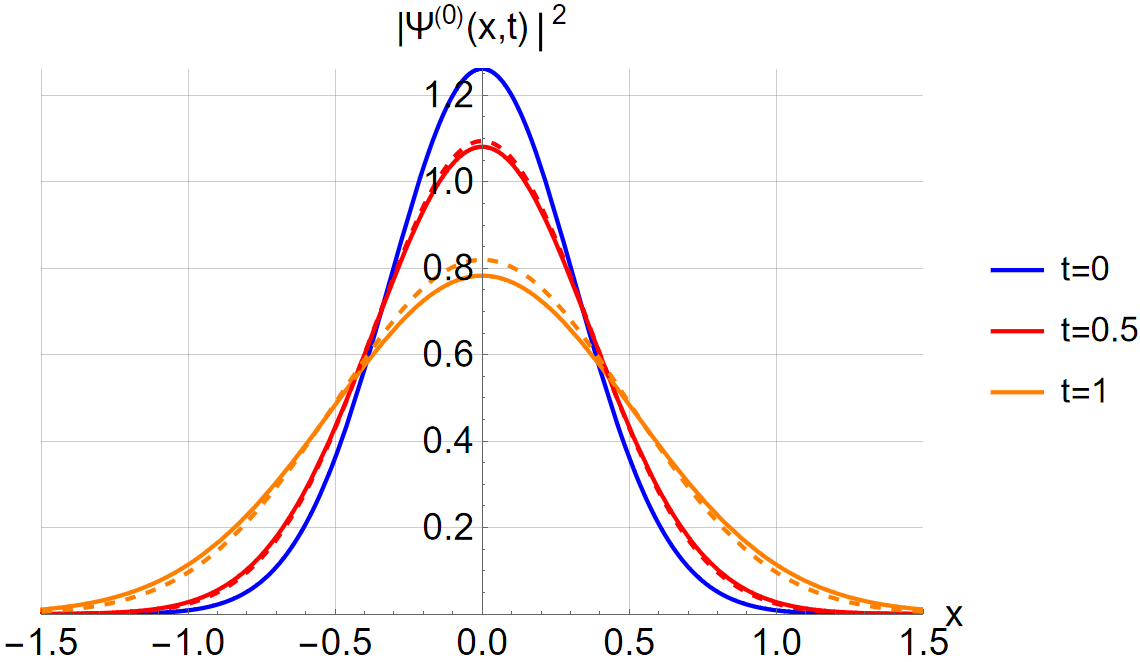} \\ c) $\Lambda=0$, $\varkappa=0.2$
  \end{minipage}
  \caption{Dependence of $|\Psi^{(0)}(x,t)|^2$ on $x$ for various $t$ and $\hbar=0.2$ (solid lines). Dashed lines are for the respective numerical solution \label{fig1}}
\end{figure}

\begin{figure}[h]
\centering\begin{minipage}[b][][b]{0.6\linewidth}\centering
    \includegraphics[width=10.5 cm]{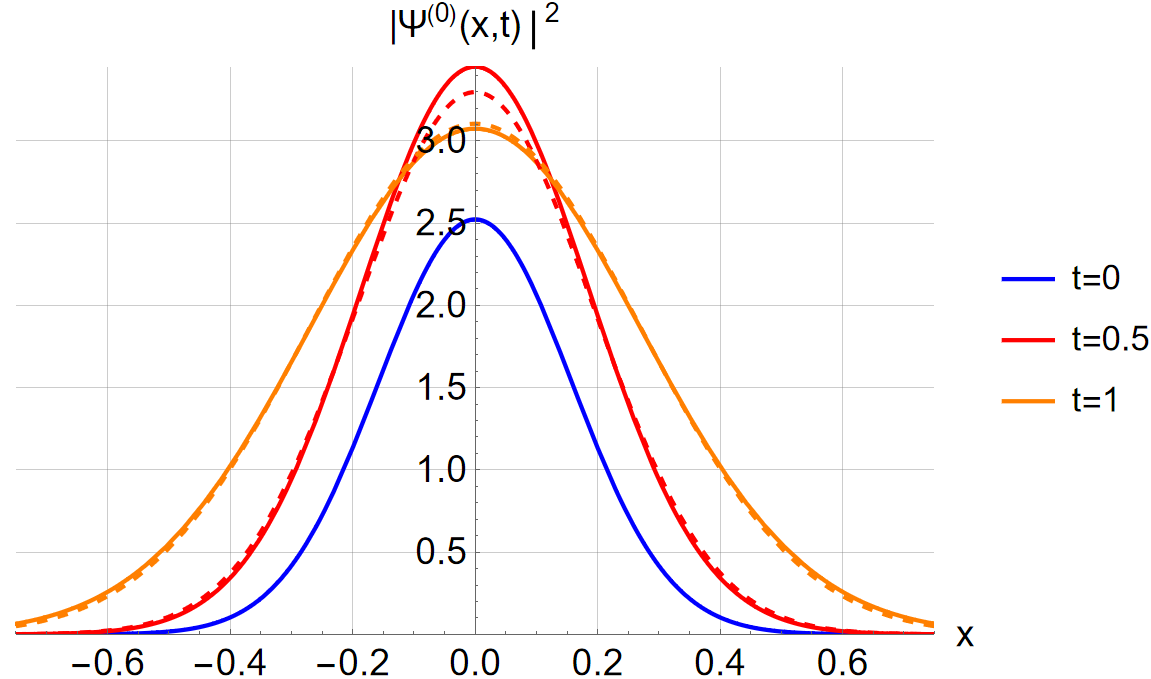} \\ a) $\Lambda=2$, $\varkappa=0.2$
  \end{minipage}\\
 \begin{minipage}[b][][b]{0.6\linewidth} \centering
    \includegraphics[width=10.5 cm]{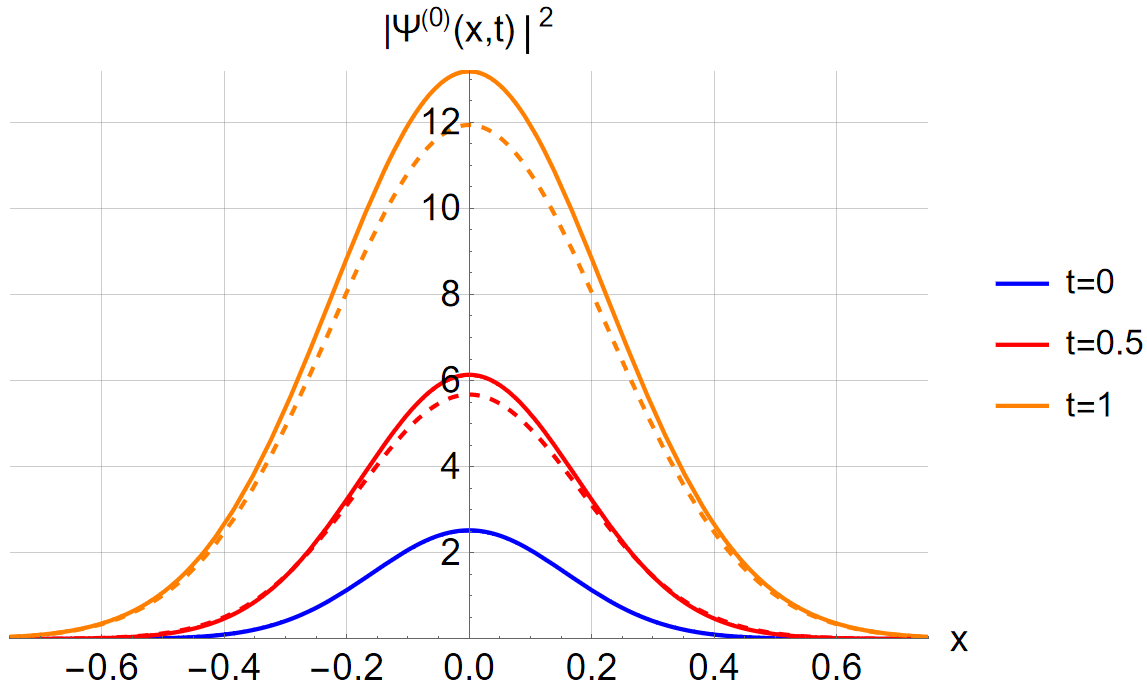} \\ b) $\Lambda=2$, $\varkappa=0$
  \end{minipage}\\
\begin{minipage}[b][][b]{0.6\linewidth} \centering
    \includegraphics[width=10.5 cm]{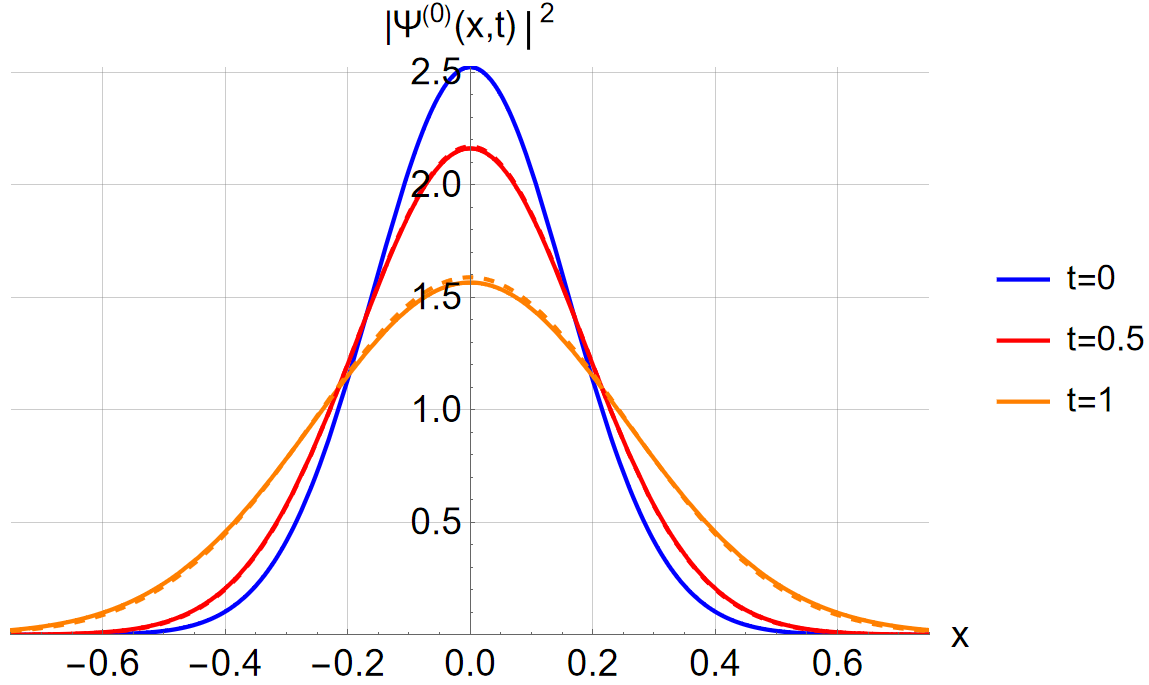} \\ c) $\Lambda=0$, $\varkappa=0.2$
  \end{minipage}
  \caption{Dependence of $|\Psi^{(0)}(x,t)|^2$ on $x$ for various $t$ and $\hbar=0.05$ (solid lines). Dashed lines are for the respective numerical solution \label{fig2}}
\end{figure}

In Fig. \ref{fig1} and \ref{fig2}, the squared absolute value of the function \eqref{prim16} is plotted, which has the meaning of the density of condensed atoms in a trap. The figures are given for $c_1=\frac{1}{2}$, $c_2=1$, $\epsilon=\frac{1}{2}$, $\gamma=1$, $N=1$, $\zeta=1$, and various values of the non-Hermiticity parameter $\Lambda$, nonlinearity parameter $\varkappa$, and small parameter $\hbar$.

We also compare the analytical asymptotic solution $\Psi^{(0)}(x,t)$ with the numerical solution to \eqref{prim1} in Fig. \ref{fig1} and \ref{fig2}. The numerical solution was obtained using the difference scheme that is common for the NLSE \cite{gpelab15}. We made the second-order spatial discretization using the method of lines with 1000 points along $x$ and the Dirichlet conditions at $x=\pm 2$. The time integration on the spatial mesh was based on the Strang--Marchuk splitting method \cite{marchuk1990} according to the following scheme:
\begin{equation}
\begin{gathered}
\pa_t\Psi(t)=\left\{\hat{A}+B[\Psi(t)]\right\}\Psi(t),\\
\left\{\hat{{\mathbb{I}}}-\dac{\tau}{4}\hat{A}\right\}\Psi(t_{n+1/2})=\left\{\hat{{\mathbb{I}}}+\dac{\tau}{4}\hat{A}\right\}\Psi(t_{n}),\\
\tilde{\Psi}(t_{n+1/2})=\dac{2+\tau B[\Psi(t_{n+1/2})]}{2-\tau B[\Psi(t_{n+1/2})]}\Psi(t_{n+1/2}),\\
\left\{\hat{{\mathbb{I}}}-\dac{\tau}{4}\hat{A}\right\}\Psi(t_{n+1})=\left\{\hat{{\mathbb{I}}}+\dac{\tau}{4}\hat{A}\right\}\tilde{\Psi}(t_{n+1/2}),
\end{gathered}
\label{scheme1}
\end{equation}
where $\hat{{\mathbb{I}}}$ is the identity operator corresponding to the identity matrix on the spatial mesh and $\tau=t_{n+1}-t_{n}$ is the time step. The operator $\hat{A}=\dac{1}{i\hbar}\left(c_1 \hat{p}^2-i\hbar\Lambda \hat{p}^2\right)$ is the differential operator from \eqref{prim1} on the spatial mesh, and the rest terms of \eqref{prim1} are included in $B[\Psi(t)]$. The initial condition was equal for both the asymptotic and numerical solutions.

In our method, $\varkappa$ and $\Lambda$ can be exactly equal to zero. In such way, we can construct asymptotic solutions to the linear Sch\"{o}dinger equation with a non-Hermitian term and to the NLSE without non-Hermitian term, respectively. However, we do not assume these parameters to be small in general. Hence, the solution behaviour can drastically change depending on $\varkappa$ and $\Lambda$. Fig. \ref{fig1} and \ref{fig2} show that the dynamics of solutions significantly changes when we put one of parameters $\varkappa$ or $\Lambda$ as well as both of them to be not equal to zero. Judging by the numerical solutions, our analytical asymptotic solutions are reasonably accurate for small $\hbar$ regardless of the presence of a non-Hermitian part.

\clearpage

\section{Conclusion}
\label{sec:con}

In this paper, we apply the semiclassical asymptotic approach to the NLSE with nonlocal cubic nonlinearity and a non-Hermite operator in the $L_2$ space \eqref{hartree1}. The non-Hermitian part of the equation accounts the environment impact on the system. We introduce the class of semiclassically concentrated functions ${\mathcal{T}}^t_\hbar$  in which we deduce evolution equations for the squared norm of the solution $\sigma_\Psi(t)$ and for the first moments $Z(t)$ describing a localization of the semiclassical solutions. These equations already provide partial but important information about the solutions to the equation. They form a dynamical system that can be considered as a nonlinear analogue of the equations of classical mechanics for a linear quantum mechanical equation. However, unlike the linear case, the dynamic moment equations depend both on the symbol of the equation operator and on the class of functions in which solutions of the equation are constructed.

To obtain the leading term of the semiclassical asymptotics of the Cauchy problem for \eqref{hartree1} explicitly within accuracy of $O(\hbar^{3/2})$, we follow the approach developed earlier for the GPE with a Hermitian equation operator (see \cite{sym2020} and references therein) in the class ${\mathcal{P}}^t_\hbar$ of trajectory concentrated functions \eqref{pth1}. The semiclassical approach required significant modification for the non-Hermitian NLSE compared to the Hermitian one due to non-conservation of the square modulus $||\Psi ||^2$ of the solution to \eqref{hartree1}. Although the general scheme of the method for constructing semiclassical asymptotics remains the same as for the nonlocal NLSE with a Hermitian operator, its implementation is subject to change. Following this scheme, we obtain a higher-order dynamical moment system (Hamilton--Ehrenfest system with dissipation) with accuracy of $O(\hbar^{3/2})$ and an associated linear Schr\"{o}dinger equation. Together with the algebraic conditions \eqref{demo3}, these equations allow us to construct a solution to the Cauchy problem in terms of the leading term of the semiclassical asymptotics with accuracy of $O(\hbar^{3/2})$. The construction of higher corrections also does not cause principal difficulties since they can be obtained using the evolution operator \eqref{ur7}, \eqref{green1}, which is given explicitly in semiclassical approximation. Also, the semiclassical symmetry operators are constructed. The general results are illustrated by the particular example of the non-Hermitian NLSE that admits explicit analytical solutions in the semiclassical approximation. The example is based on the model equation of an atom laser \cite{arecchi2000} that is a fundamentally open system.

The approach proposed is a new tool for the analytical study of open quantum systems. The semiclassical approximation was well studied for closed quantum systems using various approaches. One of their common drawbacks is that the error of semiclassical solutions (compared to the exact one) usually grows over time. However, for dissipative systems, the error can not grow indefinitely within the framework of our semiclassical approach since the dynamics of both the exact and asymptotic solutions damps over time due to the dissipation. Hence, the error of semiclassical approximation should be bounded function with respect to time for such systems. It means that the time-limited semiclassical approaches like ours can be even more natural and useful for study of open quantum systems compared to closed ones. This encourages us to further develop our method for systems with a more complex geometries of the localization domain in the future.

\section*{Acknowledgement}

The study is supported by Russian Science Foundation, project no. 23-71-01047, https://rscf.ru/en/project/23-71-01047/.

\appendix
\section{Hamilton--Ehrenfest system with dissipation of the second order}
\label{sec:app1}
Substituting the operator $A_{ij}(\hat{z},t)=\dac{1}{2}\left(\Delta\hat{z}_i\Delta\hat{z}_j+\Delta\hat{z}_j\Delta\hat{z}_i\right)$ into \eqref{mean2}, one can obtain the equation for the following matrix:
\begin{equation}
\Delta_2(t)=\Big(\langle A_{ij}(\hat{z},t)\rangle\Big)_{i,j=1}^{i,j=n}.
\label{app1}
\end{equation}
In view of estimates \eqref{estim1}, we have
\begin{equation}
\Delta_2(t)=\hbar\Delta^{(1)}(t)+\Or(\hbar^{3/2}).
\label{app2}
\end{equation}
Then, from \eqref{sig4}, \eqref{syshet5a}, we get the equation for $\Delta^{(1)}(t)$ as
\begin{equation}
\dot{\Delta}_2^{(1)}(t)=J H_{zz}\left(Z^{(0)}(t),t\right) \Delta_2^{(1)}(t) - \Delta_2^{(1)}(t) H_{zz}\left(Z^{(0)}(t),t\right) J.
\label{sge1}
\end{equation}
The equations for the first and zeroth moments \eqref{svmom1} with accuracy of $\Or(\hbar^{3/2})$ read
\begin{equation}
\begin{gathered}
\dot{Z}^{(1)}(t)=J H_{zz} \left(Z^{(0)}(t),t\right) Z^{(1)}(t)+ \dac{1}{2} J \pa_{z}\left( \Sp\left[ H_{zz}\left(Z^{(0)}(t),t\right)\cdot \Delta_2^{(1)}(t) \right]\right)-\\
-2\Lambda \breve{H}_z\left(Z^{(0)}(t),t\right) \Delta_2^{(1)}(t)+\\
+\varkappa \sigma^{(1)}(t) J W_z\left(Z^{(0)}(t),Z^{(0)}(t),t\right)+\\
+\varkappa \sigma(t) J W_{zw}\left(Z^{(0)}(t),Z^{(0)}(t),t\right)Z^{(1)}(t)+\\
+\dac{1}{2}\varkappa \sigma(t) J \pa_{z}\left( \Sp\left[ W_{ww}\left(Z^{(0)}(t),Z^{(0)}(t),t\right)\cdot \Delta_2^{(1)}(t) \right]\right).
\end{gathered}
\label{sge2}
\end{equation}

\begin{equation}
\begin{gathered}
\dot{\sigma}^{(1)}(t)=-2\Lambda \sigma(t)\langle \breve{H}_{z}\left(Z^{(0)}(t),t\right),Z^{(1)}(t)\rangle-\Lambda \sigma(t) \Sp\left[\breve{H}_{zz}\left(Z^{(0)}(t),t\right)\cdot \Delta_2^{(1)}(t)\right]-\\
-2\Lambda \sigma^{(1)}(t) H\left(Z^{(0)}(t),t\right)-\\
-2\Lambda \varkappa \left(\sigma(t)\right)^2 \langle\breve{W}_w\left(Z^{(0)}(t),Z^{(0)}(t),t\right),Z^{(1)}(t)\rangle-\\
-\Lambda \varkappa \left(\sigma(t)\right)^2 \Sp\left[ \breve{W}_{ww}\left(Z^{(0)}(t),Z^{(0)}(t),t\right)\cdot \Delta_2^{(1)}(t) \right]-\\
-2\Lambda \varkappa \sigma(t)\sigma^{(1)}(t) \breve{W}\left(Z^{(0)}(t),Z^{(0)}(t),t\right).
\end{gathered}
\label{sge4}
\end{equation}

In \eqref{sge1}, \eqref{sge2}, \eqref{sge4}, the following notations are used:
\begin{equation}
\begin{gathered}
H(z,t)=V(z,t)+\varkappa \sigma^{(0)}(t) W\left(z,Z^{(0)}(t),t\right),\\
\breve{H}(z,t)=\breve{V}(z,t)+\varkappa \sigma^{(0)}(t) \breve{W}\left(z,Z^{(0)}(t),t\right).
\end{gathered}
\label{sge3}
\end{equation}

The equations \eqref{sge1}, \eqref{sge2}, \eqref{sge4} along with \eqref{sig4}, \eqref{syshet5a} form the HESD of the second order. Note that HESD of the first order \eqref{sig4}, \eqref{syshet5a} is a closed system of nonlinear ODEs while the system \eqref{sge1}, \eqref{sge2}, \eqref{sge4} on the solutions to \eqref{sig4}, \eqref{syshet5a} is a closed system of linear homogeneous ODEs.

\section{Solution to the system \eqref{prim13}}
\label{app2}
For $c_1>0$, $c_2>0$, $\epsilon>0$, the solution to \eqref{prim13} is as follows:
\begin{equation}
\begin{gathered}
M(t)=\dac{1}{V}Q\big(\exp[2\epsilon \lambda t]\big), \qquad Q(\tau)=\begin{pmatrix} Q_1(\tau) & Q_2(\tau) \cr Q_3(\tau) & Q_4(\tau) \end{pmatrix}\\
V=G_{2,2}^{2,0}\left(k_2\middle\vert\begin{array}{c} 1-k_1,k_1+1\\ 0,1\\ \end{array} \right) \cdot \Big( c_1 c_2 N \varkappa  \, _2F_1\left(2-k_1,k_1+2;3;k_2\right)+\\
+\gamma ^2 \lambda ^2 \epsilon  \big(2 \, _2F_1\left(1-k_1,k_1+1;2;k_2\right) (\epsilon -N \varkappa )-N \varkappa  \, _2F_1\left(2-k_1,k_1+2;3;k_2\right)\big) \Big)-\\
- G_{2,2}^{2,0}\left(k_2\middle\vert\begin{array}{c} -k_1,k_1\\ 0,0\\ \end{array} \right) \cdot 2 \gamma ^2 \lambda ^2 N \epsilon  \varkappa  \, _2F_1\left(1-k_1,k_1+1;2;k_2\right), \\
k_1=\dac{\sqrt{c_1 c_2}}{\sqrt{\epsilon}\gamma \lambda}, \qquad k_2=\dac{N \varkappa}{N \varkappa- \epsilon}.
\end{gathered}
\label{vsol0}
\end{equation}
Elements of the matrix $Q(\tau)$ read
\begin{equation}
\begin{gathered}
Q_1(\tau)=G^{2,0}_{2,2}\left(\tau k_2\middle\vert\begin{array}{c} 1-k_1,k_1+1\\ 0,1\\ \end{array} \right)\cdot \Big( c_1 c_2 N \varkappa  \, _2F_1\left(2-k_1,k_1+2;3;k_2\right)+\\
+\gamma ^2 \lambda ^2 \epsilon  \left(2 \, _2F_1\left(1-k_1,k_1+1;2;k_2\right) (\epsilon -N \varkappa )-N \varkappa  \, _2F_1\left(2-k_1,k_1+2;3;k_2\right)\right)\Big)-\\
-2 \gamma ^2 \lambda ^2 N   \epsilon  \varkappa \tau \, _2F_1\left(1-k_1,k_1+1;2;\tau  k_2\right)\cdot G^{2,0}_{2,2}\left(k_2\middle\vert\begin{array}{c} 1-k_1,k_1+1\\ 0,1\\ \end{array} \right),
\end{gathered}
\label{vsol1}
\end{equation}

\begin{equation}
\begin{gathered}
Q_2(\tau)=\dac{\gamma ^2 \lambda  (N (\tau -1) \varkappa +\epsilon )}{c_2 (\epsilon -N \varkappa )}\bigg(G^{2,0}_{2,2}\left(\tau k_2\middle\vert\begin{array}{c} 1-k_1,k_1+1\\ 0,1\\ \end{array} \right)\times\\
\times \Big(c_1 c_2 N \varkappa  \, _2F_1\left(2-k_1,k_1+2;3;k_2\right)+\gamma ^2 \lambda ^2 \epsilon  \big(2 \, _2F_1\left(1-k_1,k_1+1;2;k_2\right) (\epsilon -N \varkappa )-\\
-N \varkappa  \, _2F_1\left(2-k_1,k_1+2;3;k_2\right)\big)\Big) + G^{2,0}_{2,2}\left(k_2\middle\vert\begin{array}{c} 1-k_1,k_1+1\\ 0,1\\ \end{array} \right) \times\\
\times \Big( \gamma ^2 \lambda ^2 \epsilon  \left(\text{N} \tau  \varkappa  \, _2F_1\left(2-k_1,k_1+2;3;\tau  k_2\right)-2 \, _2F_1\left(1-k_1,k_1+1;2;\tau  k_2\right) (\epsilon -\text{N} \varkappa )\right)-\\
-c_1 c_2 \text{N} \tau  \varkappa  \, _2F_1\left(2-k_1,k_1+2;3;\tau  k_2\right) \Big)\bigg),
\end{gathered}
\label{vsol2}
\end{equation}

\begin{equation}
\begin{gathered}
Q_3(\tau)=2 c_2 \lambda  N \varkappa  (\epsilon -N \varkappa ) \bigg( \tau  \, _2F_1\left(1-k_1,k_1+1;2;\tau  k_2\right)\cdot G^{2,0}_{2,2}\left(k_2\middle\vert\begin{array}{c} 1-k_1,k_1+1\\ 0,1\\ \end{array} \right)-\\
-\, _2F_1\left(1-k_1,k_1+1;2;k_2\right) \cdot G^{2,0}_{2,2}\left(\tau k_2\middle\vert\begin{array}{c} 1-k_1,k_1+1\\ 0,1\\ \end{array} \right) \bigg),
\end{gathered}
\label{vsol3}
\end{equation}

\begin{equation}
\begin{gathered}
Q_4(\tau)=\dac{N (\tau -1) \varkappa +\epsilon}{\epsilon}\bigg( G^{2,0}_{2,2}\left(k_2\middle\vert\begin{array}{c} 1-k_1,k_1+1\\ 0,1\\ \end{array} \right)\times\\
\times \Big( c_1 c_2 N \tau  \varkappa  \, _2F_1\left(2-k_1,k_1+2;3;\tau  k_2\right)+\gamma ^2 \lambda ^2 \epsilon  \big(2 \, _2F_1\left(1-k_1,k_1+1;2;\tau  k_2\right) (\epsilon -N \varkappa )-\\
-N \tau  \varkappa  \, _2F_1\left(2-k_1,k_1+2;3;\tau  k_2\right)\big) \Big)- 2 \gamma ^2 \lambda ^2 N \epsilon  \varkappa  \, _2F_1\left(1-k_1,k_1+1;2;k_2\right)\times\\
\times G^{2,0}_{2,2}\left(\tau k_2\middle\vert\begin{array}{c} 1-k_1,k_1+1\\ 0,1\\ \end{array} \right) \bigg).
\end{gathered}
\label{vsol3}
\end{equation}
Here, ${}_2F_1\left(a,b;c;z\right)$ is the hypergeometric function \cite{olver10}, and $G^{m,n}_{p,q}\left(z\middle\vert\begin{array}{c} a_1,...,a_p\\ b_1,...,b_q\\ \end{array} \right)$ is the Meijer G-function \cite{karp22}.

\bibliography{lit1}

\begin{thebibliography}{10}
\newcommand{\enquote}[1]{``#1''}
\expandafter\ifx\csname url\endcsname\relax
  \def\url#1{\texttt{#1}}\fi
\expandafter\ifx\csname urlprefix\endcsname\relax\def\urlprefix{URL }\fi
\providecommand{\eprint}[2][]{\url{#2}}

\bibitem{lederer2008}
F.~Lederer, G.~Stegeman, D.~Christodoulides, G.~Assanto, M.~Segev, and
  Y.~Silberberg, \enquote{Discrete solitons in optics,} Physics Reports-review
  Section of Physics Letters \textbf{463}, 1--126 (2008).

\bibitem{pitaevskii1999}
F.~Dalfovo, S.~Giorgini, L.~P. Pitaevskii, and S.~Stringari, \enquote{Theory of
  Bose-Einstein condensation in trapped gases,} Reviews of Modern Physics
  \textbf{71}(3), 463--512 (1999).

\bibitem{leggett2001}
A.~Leggett, \enquote{Bose-Einstein condensation in the alkali gases: Some
  fundamental concepts,} Reviews of Modern Physics \textbf{73}, 307--356
  (2001).

\bibitem{saffman1993}
P.~G. Saffman, \emph{Vortex Dynamics}, Cambridge Monographs on Mechanics
  (Cambridge University Press, 1993).

\bibitem{fetter01}
A.~L. Fetter and A.~A. Svidzinsky, \enquote{Vortices in a trapped dilute
  Bose-Einstein condensate,} Journal of Physics Condensed Matter
  \textbf{13}(12), 135--194 (2001).

\bibitem{hasegawa1973}
A.~Hasegawa and F.~Tappert, \enquote{Transmission of stationary nonlinear
  optical pulses in dispersive dielectric fibers. II. Normal dispersion,}
  Applied Physics Letters \textbf{23}, 171 -- 172 (1973).

\bibitem{mollenauer1980}
L.~F. Mollenauer, R.~H. Stolen, and J.~P. Gordon, \enquote{Experimental
  Observation of Picosecond Pulse Narrowing and Solitons in Optical Fibers,}
  Phys. Rev. Lett. \textbf{45}, 1095--1098 (1980).

\bibitem{zakharov72}
V.~Zakharov and A.~Shabat, \enquote{Exact Theory of Two-Dimensional
  Self-Focusing and One-Dimensional Self-Modulation of Waves in Nonlinear
  Media,} Soviet Physics Journal of Experimental and Theoretical Physics
  Letters \textbf{34}, 62--69 (1972).

\bibitem{arecchi2000}
F.~Arecchi, J.~Bragard, and L.~Castellano, \enquote{Dissipative dynamics of an
  open Bose Einstein condensate,} Optics Communications \textbf{179}(1),
  149--156 (2000).

\bibitem{aranson2002}
I.~S. Aranson and L.~Kramer, \enquote{The world of the complex Ginzburg-Landau
  equation,} Rev. Mod. Phys. \textbf{74}, 99--143 (2002).

\bibitem{brazhnyi2004}
V.~Brazhnyi and V.~Konotop, \enquote{Theory of nonlinear matter waves in
  optical lattices,} Modern Physics Letters B \textbf{18}(14), 627--651 (2004).

\bibitem{abdullaev2010}
F.~Abdullaev, V.~Konotop, M.~Salerno, and A.~Yulin, \enquote{Dissipative
  periodic waves, solitons, and breathers of the nonlinear Schrodinger equation
  with complex potentials,} Physical review. E, Statistical, nonlinear, and
  soft matter physics \textbf{82}, 056606 (2010).

\bibitem{sels2020}
D.~Sels and E.~Demler, \enquote{Thermal radiation and dissipative phase
  transition in a BEC with local loss,} Annals of Physics \textbf{412}, 168021
  (2020).

\bibitem{haus1984}
H.~Haus, \emph{Waves and Fields in Optoelectronics}, Prentice-Hall series in
  solid state physical electronics (Prentice-Hall, 1984).

\bibitem{aleksic21}
B.~Aleksic, L.~Uvarova, and N.~Aleksić, \enquote{Dissipative structures in the
  resonant interaction of laser radiation with nonlinear dispersive medium,}
  Optical and Quantum Electronics \textbf{53} (2021).

\bibitem{aleksic20}
B.~Aleksic, L.~Uvarova, N.~Aleksić, and M.~Belić, \enquote{Cubic quintic
  Ginzburg Landau equation as a model for resonant interaction of EM field with
  nonlinear media,} Optical and Quantum Electronics \textbf{52} (2020).

\bibitem{aleksic14}
B.~N. Aleksi\'{c}, N.~B. Aleksi\'{c}, M.~S. Petrovi\'{c}, A.~I. Strini\'{c},
  and M.~R. Beli\'{c}, \enquote{Variational and accessible soliton
  approximations to multidimensional solitons in highly nonlocal nonlinear
  media,} Opt. Express \textbf{22}(26), 31842--31852 (2014).

\bibitem{baranov2008}
M.~A. Baranov, \enquote{Theoretical progress in many-body physics with
  ultracold dipolar gases,} Physics Reports \textbf{464}(3), 71--111 (2008).

\bibitem{malomed2009}
J.~Cuevas, B.~A. Malomed, P.~G. Kevrekidis, and D.~J. Frantzeskakis,
  \enquote{Solitons in quasi-one-dimensional Bose-Einstein condensates with
  competing dipolar and local interactions,} Physical Review A - Atomic,
  Molecular, and Optical Physics \textbf{79}(5) (2009).

\bibitem{klaus2022}
L.~Klaus, T.~Bland, E.~Poli, C.~Politi, G.~Lamporesi, E.~Casotti, R.~Bisset,
  M.~Mark, and F.~Ferlaino, \enquote{Observation of vortices and vortex stripes
  in a dipolar condensate,} Nature Physics \textbf{18}, 1--6 (2022).

\bibitem{zhao2021}
Q.~Zhao, \enquote{Effects of Dipole-Dipole Interaction on Vortex Motion in
  Bose-Einstein Condensates,} Journal of Low Temperature Physics \textbf{204},
  1--11 (2021).

\bibitem{curtis2012}
C.~W. Curtis, \enquote{On nonlocal Gross-Pitaevskii equations with periodic
  potentials,} Journal of Mathematical Physics \textbf{53}(7), 073709 (2012).

\bibitem{Maslov2}
V.~Maslov, \emph{The Complex WKB Method for Nonlinear Equations. I. Linear
  Theory} (Birkhauser Verlag, Basel, 1994).

\bibitem{dobrokhotov2022}
S.~Dobrokhotov, D.~Minenkov, and V.~Nazaikinskii, \enquote{Asymptotic Solutions
  of the Cauchy Problem for the Nonlinear Shallow Water Equations in a Basin
  with a Gently Sloping Beach,} Russian Journal of Mathematical Physics
  \textbf{29}, 28--36 (2022).

\bibitem{pereskokov2022}
A.~Pereskokov, \enquote{Semiclassical Asymptotics of the Spectrum of a
  Two-Dimensional Hartree Type Operator Near Boundaries of Spectral Clusters,}
  Journal of Mathematical Sciences \textbf{264} (2022).

\bibitem{shapovalov:BTS1}
V.~V. Belov, A.~Y. Trifonov, and A.~V. Shapovalov, \enquote{The
  trajectory-coherent approximation and the system of moments for the hartree
  type equation,} International Journal of Mathematics and Mathematical
  Sciences \textbf{32}(6), 325--370 (2002).

\bibitem{sym2020}
A.~V. Shapovalov, A.~E. Kulagin, and A.~Y. Trifonov, \enquote{The
  Gross--Pitaevskii equation with a nonlocal interaction in a semiclassical
  approximation on a curve,} Symmetry \textbf{12}(2), 201 (2020).

\bibitem{kulagin2021}
A.~E. Kulagin, A.~V. Shapovalov, and A.~Y. Trifonov, \enquote{Semiclassical
  spectral series localized on a curve for the Gross--Pitaevskii equation with
  a nonlocal interaction,} Symmetry \textbf{13}(7), 1289 (2021).

\bibitem{fkppshap2018}
A.~V. Shapovalov and A.~Y. Trifonov, \enquote{An application of the Maslov
  complex germ method to the one-dimensional nonlocal Fisher-KPP equation,}
  International Journal of Geometric Methods in Modern Physics \textbf{15}(6)
  (2018).

\bibitem{shapkul21}
A.~Shapovalov and A.~Kulagin, \enquote{Semiclassical approach to the nonlocal
  kinetic model of metal vapor active media,} Mathematics \textbf{9}(23), 2995
  (2021).

\bibitem{shapkul22}
A.~Shapovalov, A.~Kulagin, and S.~Siniukov, \enquote{Family of Asymptotic
  Solutions to the Two-Dimensional Kinetic Equation with a Nonlocal Cubic
  Nonlinearity,} Symmetry \textbf{14}(3), 577 (2022).

\bibitem{nath2017}
D.~Nath and P.~Roy, \enquote{Nonlinear Schrodinger Equation with Complex
  Supersymmetric Potentials,} Physics of Particles and Nuclei Letters
  \textbf{14}, 347--356 (2017).

\bibitem{Maslov1}
V.~Maslov, \emph{Operational Methods} (Mir Publishers, Moscow, 1976).

\bibitem{multiind}
M.~V. Karasev, \enquote{Weyl and ordered calculus of noncommuting operators,}
  Mathematical Notes of the Academy of Sciences of the USSR \textbf{26}(6),
  945--958 (1979).

\bibitem{bojowald05}
M.~Bojowald and A.~Skirzewski, \enquote{Effective Equations of Motion for
  Quantum Systems,} Reviews in Mathematical Physics \textbf{18}(07), 713--745
  (2005).

\bibitem{bagrov1}
V.~G. Bagrov, V.~V. Belov, and A.~Y. Trifonov, \enquote{Semiclassical
  trajectory-coherent approximation in quantum mechanics I. High-order
  corrections to multidimensional time-dependent equations of Schrödinger
  type,} Annals of Physics \textbf{246}(2), 231--290 (1996).

\bibitem{schweber61}
S.~Schweber, \emph{An introduction to relativistic Quantum Field Theory} (Row,
  Peterson and Co., Evanston, 1963).

\bibitem{kneer98}
B.~Kneer, T.~Wong, K.~Vogel, W.~P. Schleich, and D.~F. Walls, \enquote{Generic
  model of an atom laser,} Phys. Rev. A \textbf{58}, 4841--4853 (1998).

\bibitem{gpelab15}
X.~Antoine and R.~Duboscq, \enquote{GPELab, a Matlab Toolbox to solve
  Gross-Pitaevskii Equations II: dynamics and stochastic simulations,} Computer
  Physics Communications  (2015).

\bibitem{marchuk1990}
G.~Marchuk, \enquote{Splitting and alternating direction methods,} vol.~1 of
  \emph{Handbook of Numerical Analysis}, pp. 197--462 (Elsevier, 1990).

\bibitem{olver10}
F.~W.~J. Olver, D.~W. Lozier, R.~F. Boisvert, and C.~W. Clark, eds., \emph{NIST
  handbook of mathematical functions} (Cambridge University Press, Cambridge,
  2010).

\bibitem{karp22}
D.~Karp and E.~Prilepkina, \enquote{On Meijer's G function G m,n p,p for m + n
  = p,} Integral Transforms and Special Functions \textbf{34}, 1--17 (2022).

\end{thebibliography}

\end{document}